\documentclass{kapproc}
\usepackage[dvips]{graphicx}
\upperandlowercase
\begin{document}
\articletitle{Outstanding Problems in Particle Astrophysics$^{\ast}$}
\author{Thomas K. Gaisser$^{\dagger}$}
\affil{$^{\dagger}$Bartol Research Institute, University of
Delaware, Newark, DE 19716, USA}
\email{gaisser@bartol.udel.edu}
\vskip0.1in
{\footnotesize
\noindent
{$^{\ast}$Work supported in part by the U.S. Department of 
Energy under DE-FG02 91ER40626}
}

\begin{abstract}
The general features of the cosmic-ray spectrum have been known for a long time.
Although the basic approaches to understanding cosmic-ray propagation and acceleration
have also been well understood for many years, there are several questions of great
interest that motivate the current intense experimental activity in the field.  
If the energy-dependence of the secondary to primary ratio
of galactic cosmic rays is as steep as observed,
why is the flux of PeV particles so nearly isotropic?
Can all antiprotons and positrons be explained as secondaries
or is there some contribution from exotic sources? 
What is the maximum energy of cosmic accelerators?
Is the "knee" of the cosmic-ray spectrum an effect of propagation
or does it perhaps reflect the upper limit of galactic acceleration processes?
Are gamma-ray burst sources (GRBs) and/or active galactic nuclei (AGN)
accelerators of ultra-high-energy cosmic rays (UHECR) as well 
as sources of high-energy photons?
Are GRBs and/or AGNs also sources of high-energy neutrinos?
If there are indeed particles with energies greater than the cutoff expected
from propagation through the microwave background radiation,
what are their sources?
The purpose of this lecture is to introduce the main topics of the School
and to relate the theoretical questions to the experiments that can answer them.
\end{abstract}
\begin{figure}[htb]
\includegraphics[width=5.0 truein]{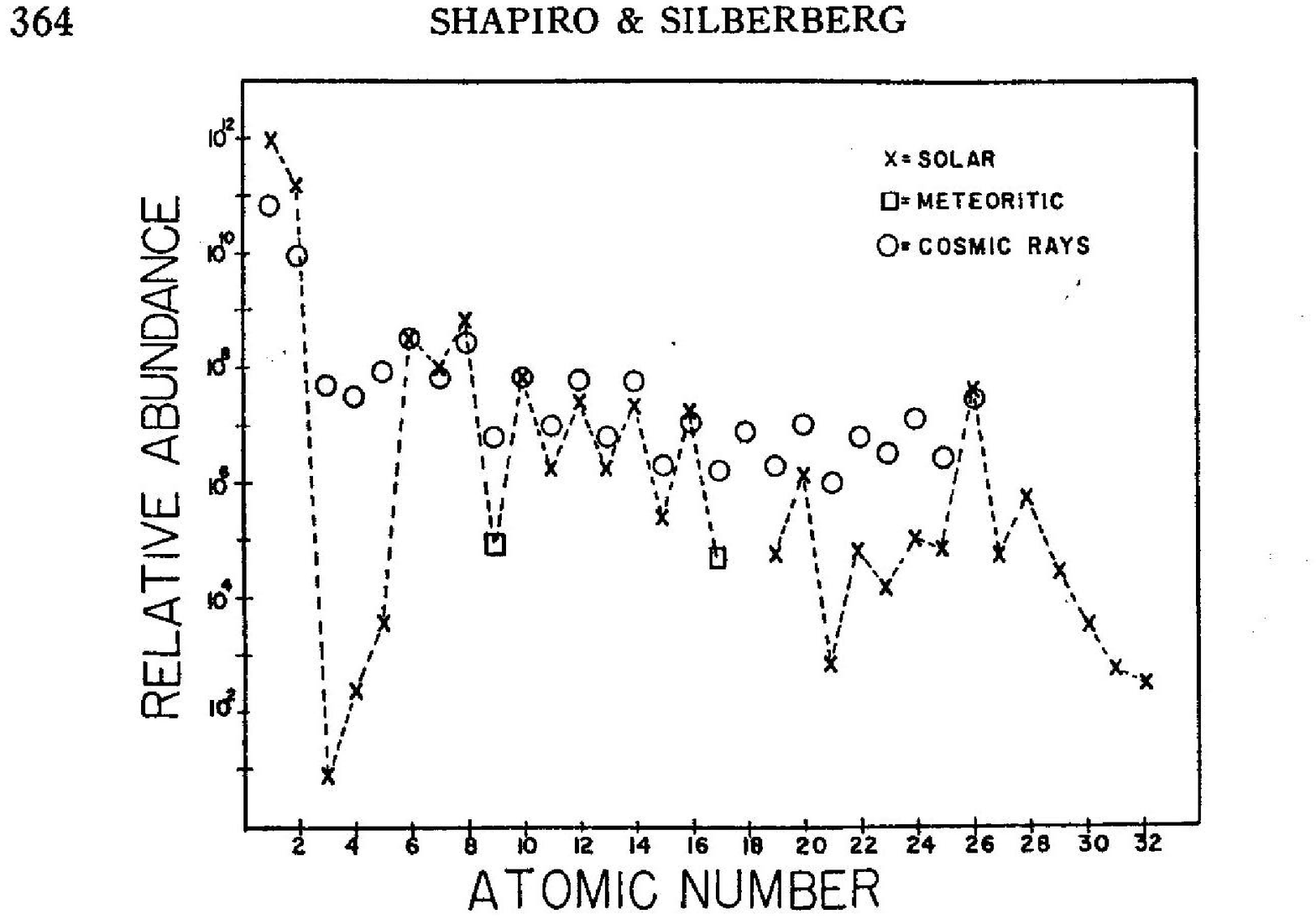}
\caption{Comparison of cosmic-ray abundances with abundances of
nuclei in the solar system~(\protect\cite{ShapSilb}).
}
\label{shapiro}
\end{figure}

\section{Introduction}
It is appropriate to begin this lecture with a diagram from the
review of \cite{ShapSilb},
 which compares the abundances
of elements in the cosmic radiation with solar system abundances.
This classic measurement is one of the foundations of cosmic-ray physics.
The elements lithium, beryllium and boron are quite abundant among
cosmic rays even though they constitute only a tiny fraction of the material
in the solar system and the interstellar medium.  
This fact is understood largely as the result of
spallation of the ``primary'' nuclei carbon and oxygen during their
propagation in the interstellar medium.  The idea is that
the protons and nuclei are accelerated from the interstellar medium
and/or from the gas in or near their sources.  The composition of
the ``primary'' nuclei (defined as those initially accelerated and
injected into the interstellar medium with high energy) reflects
the combination of nuclei present in the material that is accelerated,
which contains negligible amounts of ``secondary" elements such as
Li, Be, B.  These secondary nuclei are fragmentation products of
heavier primary nuclei.

In this picture, the amount of secondary nuclei is a measure of the
characteristic time for propagation of cosmic rays before they escape
from the galaxy into inter-galactic space.  A simplified version of the
diffusion equation that relates the observed abundances and spectra to initial
values is
\begin{equation}
{N_i(E)\over \tau_{esc}(E)}\,=\,Q_i(E)
-\left( \beta c n_H \sigma_i + {1\over \gamma\tau_i}\right)N_i(E)
+\beta c n_H\sum_{k\ge i}\sigma_{k\rightarrow i}\,N_k(E).
\label{propagation}
\end{equation}

Here $N_i(E)$ is the spatial density of cosmic-ray nuclei
of mass $i$, and $n_H$ is the number density of target nuclei (mostly hydrogen) 
in the interstellar medium, $Q_i(E)$ is the number of primary nuclei
of type $i$ accelerated per cm$^3$ per second, 
and $\sigma_i$ and $\sigma_{k\rightarrow i}$
are respectively the total and partial cross sections for interactions
of cosmic-ray nuclei with the gas in the interstellar medium.  The
second term on the r.h.s. of Eq.~\ref{propagation} represents losses due
to interactions with cross section $\sigma_i$
and decay for unstable nuclei with lifetime $\tau_i$.
  The energy per nucleon, $E$,
remains constant to a good approximation in the transition from parent
nuclei to nuclear spallation products, which move with velocity $\beta c$
and Lorentz factor $\gamma\,=\,E/m_p$.

A crude first estimate of the characteristic diffusion time $\tau_{esc}$ can
be made by neglecting propagation losses for a primary nucleus $P$ and assuming
that $Q_S=0$ for a secondary nucleus $S$.  If we also neglect collision losses
by the secondary nucleus after it is produced, then the solution of 
Eq.~\ref{propagation} is
\begin{equation}
n_H\,\tau_{esc}\;=\;{1\over \beta c\sigma_{P\rightarrow S}}\,{N_S\over N_P}.
\label{time}
\end{equation}
Since the density in the disk of the galaxy is of order one particle per cm$^3$
and the typical partial cross section for a light nucleus will be of order 
$100$~mb, the scale for the characteristic time is $\sim10^7$~years.
The full analysis requires self-consistent solution of the coupled equations
for all species accounting for all loss terms, as described in John Wefel's 
lecture in this volume~(\cite{Wefel}).  A nice overview of propagation models
is given by \cite{Jonesetal}.

The thickness of the disk of the galaxy is of order 300 pc = 1000 light years,
which is much shorter than the characteristic propagation time of 10 million years.
The explanation is that the charged particles are trapped in the turbulent magnetized
plasma of the interstellar medium and only diffuse slowly away from the disk, which is
assumed to be where the sources are located.  Measurements of the ratio of unstable
to stable secondary nuclei (especially $^{10}Be/^{9}Be$) are used to determine
$\tau_{esc}$ independently of the product $n_H\,\tau_{esc}$ and hence to constrain
further the models of cosmic-ray propagation.

Another important fact is that the ratio $N_S/N_P$ is observed to decrease with 
energy.  From Eq.~\ref{time} this implies that $\tau_{esc}$ also decreases.
Simple power-law fits to ratios like B/C give
\begin{equation}
\tau_{esc}\;\propto\;E^{-\delta},
\label{Edependence}
\end{equation}
with $\delta\approx 0.6$.
This behavior has important consequences for the source spectrum.  To see
this, consider an abundant, light primary nucleus
such as hydrogen or helium.  They are sufficiently abundant so that feed-down
from heavier nuclei can be neglected and their cross sections are small
enough so that energy losses in the interstellar medium can also be neglected
in a first approximation.
Then Eq.~\ref{propagation} reduces to
\begin{equation}
Q(E)\;=\;N(E)/ \tau_{esc}\;\approx\;N(E)\times E^\delta.
\label{source}
\end{equation}
The local energy-density spectrum of cosmic rays is related to the
observed flux $\phi$(particles per cm$^2$ per GeV per second per steradian) by 
\begin{equation}
N(E)\;=\;{4\pi\over c}\,\phi(E).
\label{fluxdef}
\end{equation}
Since $\phi(E)\approx E^{-\alpha}$ with $\alpha\approx 2.7$ the inference
is that the cosmic accelerators are characterized by a power law
with spectral index $\alpha_S\approx 2.1$.  This value is close
to the spectral index for first order acceleration by strong
shocks in the test-particle approximation~(\cite{Ostrowski}).

If the time for diffusion out of the galaxy
continues to be described by $\delta \approx 0.6$ to very high energy
there is a problem: as $c\tau_{esc}$
decreases and approaches galactic scales, the cosmic-ray fluxes should become
significantly anisotropic, which is not observed.  One possibility is that the
observed energy-dependence at low energy is due 
to a combination of ``reacceleration''~(\cite{reacceleration1,reacceleration2})
by weak shocks in the interstellar medium after initial acceleration by strong 
shocks.  In such models, the high-energy behavior of diffusion is typically
described by a slower energy dependence with $\delta\approx 0.3$.  If so, the
source spectral index would be steeper, approximately $\alpha\,_S=2.4$.  
In any case, in more realistic non-linear treatments of acceleration by
strong shocks, the spectrum has some curvature, being steeper at low energy
and harder at the high energy end of the spectrum~(\cite{nonlinear}).  
What we observe may be
some kind of average over many sources, each of which is somewhat different
in shape and maximum energy.

The assumption underlying the discussion above is that the sources accelerating
cosmic rays are in the disk of the galaxy and that the energy density in cosmic
rays observed locally is typical of other regions of the galactic disk.
If so, the total power $P_{CR}$ required to maintain the cosmic radiation in
equilibrium may be obtained by integrating Eq.~\ref{source} over energy and
space.  The result is
\begin{equation}
P_{CR}\;=\;\int\,{\rm d}^3x\,\int\,Q(E)\,{\rm d}E\;=\;
V_G\,{4\pi\over c}\,\int\,\phi(E)\,/\,\tau_{esc}(E)\,{\rm d}E.
\label{power}
\end{equation}
Using the observed spectrum and the value of $\tau_{esc}$ explained above, 
one finds numerically
\begin{equation}
P_{CR}\;\sim\;10^{41}\,{\rm erg/s}.
\label{Galpower}
\end{equation}
The kinetic energy of an expanding supernova remnant is initially
of order $10^{51}$~erg, excluding the neutrinos, which carry
away most of the energy released but do not disturb the
interstellar medium.  There are perhaps 3 supernovae per century,
which gives $P_{SN}\sim 10^{42}$~erg/s as an estimate of the power which
is dissipated in the interstellar medium by means of strong
shocks driven by supernova ejecta.  
These are just the ingredients needed for acceleration
of galactic cosmic rays.  I return to the subject of cosmic-ray
acceleration in \S3 below.

\section{Secondary cosmic-rays}
Depending on the context, the term {\it secondary cosmic rays} can refer
either to particles produced by interactions
of primary cosmic rays with the interstellar gas 
or to particles produced
by interactions of cosmic rays in the Earth's atmosphere.  The production
mechanisms are similar, and there are some common features.
 
We have already seen one example, the secondary nuclei produced by occasional
interactions of primary cosmic-ray nuclei 
during their propagation in the interstellar medium.
In that case, to a good approximation, the energy per nucleon of the
secondary nucleus is the same as that of the parent nucleus.  The reason
is that the nuclear fragments are only spectators to any production of secondary
pions that may occur in the collisions.  In general, however, the production
term (last term of Eq.~\ref{propagation}) will involve an integration over
the energy of the parent particle.  The main process to consider is 
production of
pions by interaction of protons with a target nucleus.
The production spectrum of the pions is
\begin{equation}
{\rm d}\phi_\pi(E_\pi)\;=\;{{\rm d}X\over\lambda_p}\,\int_{E_\pi}^\infty\,
\phi_p(E_p)\,{{\rm d}n_\pi(E_\pi,E_p)\over {\rm d}E_\pi}\,{\rm d}E_p.
\label{production}
\end{equation}
Here $\lambda_p = A\,m_p/\sigma_{pA}^{\rm inel}$ is the interaction length
of protons in a medium consisting of nuclei of mass $A$, and d$X=\rho\,{\rm d}\ell$
is the differential element of mass traversed in distance d$\ell$
in a medium of density $\rho$.
It is often a useful approximation at high energy (i.e. energy $\gg$ particle masses)
to assume a scaling form for the dimensionless production spectrum:
\begin{equation}
E_p{{\rm d}n_\pi(E_\pi,E_p)\over {\rm d}E_\pi}\;=\;{{\rm d}n_\pi(\xi)\over {\rm d}\xi}.
\label{scaling}
\end{equation}
The scaling variable is $\xi = E_\pi/E_p$, and Eq.~\ref{production} becomes
\begin{equation}
{\rm d}\phi_\pi(E_\pi)
\;=\;{{\rm d}X\over\lambda_p}\,\int_0^1\,
\phi_p\left({E_\pi\over\xi}\right){{\rm d}n_\pi(\xi)\over {\rm d}\xi}
{{\rm d}\xi\over\xi}\rightarrow{{\rm d}X\over\lambda_p}\,K\,E_\pi^{-\alpha}
\,Z_{p\rightarrow\pi}(\alpha).
\label{scaling2}
\end{equation}
The last step on the r.h.s. of Eq.~\ref{scaling2} follows when
the parent spectrum is a power law in energy ($\phi_p(E)=K\,E^{-\alpha}$).
In that case, in the high-energy scaling approximation
\begin{equation}
{\rm d}\phi_\pi(E_\pi)\rightarrow 
{{\rm d}X\over\lambda_p}Z_{p\rightarrow\pi}\times\phi_p(E_\pi),
\label{scaling3}
\end{equation}
i.e. the energy spectrum of secondaries has the same power behavior as 
the primaries scaled down by a factor
\begin{equation}
Z_{p\rightarrow\pi}(\alpha)\;=\;\int_0^1\,\xi^{\alpha-1}{{\rm d}n_\pi
\over{\rm d}\xi}\,{\rm d}\xi.
\label{Zfactor}
\end{equation}
The spectrum-weighted moment $Z_{p\rightarrow\pi}(\alpha)$
depends both on the physics of production of the secondary pion
and on the value of the differential spectral index $\alpha$.

\subsection{Galactic secondaries}

\subsubsection{Diffuse gamma-rays}
Gamma-ray emission from the disk of the galaxy is a powerful
probe of the model of cosmic-ray origin and propagation as well
as of the structure of the galaxy.  Unlike charged cosmic rays,
secondary photons propagate in straight lines.  Since the galaxy
is transparent for $\gamma$-rays of most energies, it is possible
to search for concentrations of primary cosmic-ray
activity from the map of the $\gamma$-ray sky 
after subtracting point sources.  For example, if cosmic-ray
acceleration is correlated with regions of higher density
such as star-forming regions where supernovae are more frequent,
then one would expect a quadratic enhancement of secondary
production because of the spatial correlation between
primary flux and target density.

The baseline calculation is to assume that the intensity
observed locally at Earth is representative of the distribution
everywhere in the disk of the galaxy.  One can then look for
interesting variation superimposed on this baseline.
Following the analysis of 
Eqs.~[\ref{scaling},\ref{scaling2},\ref{scaling3},\ref{Zfactor}],
the average number of neutral pions produced per GeV per
unit volume in the interstellar medium is
\begin{equation}
q_\pi\;=\;4\pi\,n_H\,\sigma_{pH}^{\rm inel}\,Z_{p\rightarrow\pi^0}\,\phi_p(E_\pi).
\label{pi0prod}
\end{equation}

Next this expression has to be convolved with
the distribution of photons produced in $\pi^0\rightarrow\,\gamma\,\gamma$.
In the rest frame of the parent pion, each photon has $E_\gamma = m_\pi/2 = 70$~MeV.
The angular distribution is isotropic, so
\begin{equation}{{\rm d}n_\gamma\over{\rm d}\Omega^*}
\;=\;{1\over 2\pi}\,{{\rm d}n_\gamma\over{\rm d}\cos\theta^*}\;=\;{1\over 2\pi} ,
\label{pi0decay}
\end{equation}
where $\theta^*$ is the polar angle of the photon
along the direction of motion of the parent pion
but evaluated in the rest frame of the pion.  
For decay in flight of a pion with Lorentz factor $\gamma$
and velocity $\beta c$, the energy of each of the
resulting photons is
\begin{equation}
E_\gamma\;=\;\gamma {m_\pi\over 2}\,(1\,+\,\beta\cos\theta^*)
\label{Egamma}
\end{equation}
with $\cos\theta_1^*\,=\,-\cos\theta_2^*$ for the two photons.
Changing variables in Eq.~\ref{pi0decay} then gives
\begin{equation}
{{\rm d}n_\gamma\over{\rm d}E_\gamma}\;=\;{2\over\beta\gamma m_\pi}.
\label{pi0}
\end{equation}
For $E_\pi > E_\gamma\gg m_\pi/2$
the convolution of the distribution~(\ref{pi0}) with the production
spectrum of neutral pions~(\ref{pi0prod}) gives
\begin{equation}
q_\gamma(E_\gamma)\;\approx\;4\pi\,n_H\,\sigma_{pH}^{inel}\times
{2\over\alpha}\,Z_{p\rightarrow\pi^0}(\alpha)
\,\phi_p(E_\gamma).
\label{pigamma}
\end{equation}

Numerically, in the approximation of uniform cosmic-ray density and
uniform gas density in the interstellar medium, the observed 
gamma-ray flux would be
\begin{equation}
{\phi_\gamma(E_\gamma)\over\phi_p(E_\gamma)}\;\approx\;
3\times 10^{-6}\,\left({n_H\over{\rm cm}^3}\right)
\,\left({r_{max}(b,\phi_\ell)\over 1 kpc}\right).
\label{numerical}
\end{equation}
Here $r_{max}(b,\phi_\ell)$ is the distance in a direction
$\{b,\phi_\ell\}$ to the effective edge of
the galactic disk, where $b$ and $\phi_\ell$ are galactic latitude
and longitude.  The effective distance is defined with respect to
an equivalent disk of uniform density.
Eq.~\ref{numerical} compares well in order of magnitude with
the measured intensity of GeV photons by Egret~(\cite{Egret}).
The derivation of Eq.~\ref{numerical} is grossly oversimplified
compared to the actual model calculation made in the paper of~\cite{Egret} 
of the diffuse, galactic gamma-radiation.  The reader is urged
to consult that paper to understand the impressive level of
detail at which the data are understood.
It is also interesting to compare Eq.~\ref{numerical} to the TeV diffuse
flux measured by Milagro~(\cite{Goodman}).

The implication of Eq.~\ref{pigamma} is that for $E_\gamma\gg 70$~MeV
the diffuse gamma-ray spectrum should have the same power law
behavior as the proton spectrum, $\alpha\approx 2.7$.
What is observed, however, is that the spectrum of gamma-rays from
the inner galaxy is harder than this, having a power-law
behavior of approximately $E_\gamma^{-2.4}$~(\cite{Egret}).  This is currently
not fully understood.  One possibility is that the cosmic-ray
spectrum producing the gamma rays is harder than observed locally
near Earth~(\cite{Egret}).

Cosmic-ray electrons also contribute to the
diffuse gamma-radiation by bremsstrahlung and by inverse Compton
scattering.  Fitting the observed spectrum requires a complete
model of propagation that includes all  
contributions~(\cite{Egret}).
The distinguishing feature of $\pi^0$-decay photons is
a kinematic peak at $E_\gamma\,=\,m_\pi/2$.  The origin of this
feature can be seen in Eq.~\ref{Egamma} from
which the limits on $E_\gamma$ for any given Lorentz factor of
the parent pion are
$$\sqrt{1-\beta\over 1+\beta}{m_\pi\over 2} < 
E_\gamma < \sqrt{1+\beta\over 1-\beta}{m_\pi\over 2}.$$
The distribution $dn_\gamma/dE_\gamma$ is flat between these limits for each 
$\gamma$ and is always centered around $\ln(m_\pi/2)$ when plotted
as a function $\ln(E_\gamma)$.  The individual contributions for
parent pions of various energies always overlap at $\ln(E_\gamma)=m_\pi/2$,
so the full distribution from any spectrum always peaks at this
value (\cite{Stecker}).

\subsubsection{Antiprotons and positrons}

Antiprotons and positrons are of special interest because
an excess over what is expected from production by protons
during propagation could reflect an exotic process such as
evaporation of primordial black holes or decay of exotic
relic particles~(\cite{exotica}).  At a more practical level,
they are important because they are secondaries of the
dominant proton component of the cosmic radiation.  As a
consequence their spectra and abundances provide an
independent constraint on models of cosmic-ray 
propagation (~\cite{MoskStrongpbar}). 

Secondary antiprotons have a kinematic feature 
analogous to that in $\pi^0$-decay gamma rays but at a
higher energy related to the nucleon mass.  In this
case the feature is related to the high threshold
for production of a nucleon-antinucleon pair in
a proton-proton collision.  This kinematic feature is
observed in the data~(\cite{BESSpbar}), and suggests that an exotic
component of antiprotons is not required.  Antiproton
fluxes are consistent with the basic model of cosmic-ray
propagation described in the Introduction.

Positrons are produced in the chain $$
p\rightarrow\pi^+\rightarrow\mu^+\rightarrow e^+.$$
Secondary electrons are produced in the charge conjugate
process, but their number in the GeV range is an order of magnitude
lower than primary electrons (i.e. electrons accelerated
as cosmic rays).
Because of radiative processes, the spectra of positrons
and electrons are more complex to interpret than
high-energy secondary $\gamma$-rays~(\cite{MoskStronge}).  
The measured intensity of positrons appears to be consistent
with secondary origin (\cite{HEAT}).
 
\subsubsection{$\gamma$-rays and $\nu$ from young supernova remnants}
The same equations that govern production of secondaries
in the interstellar medium also apply to production 
in gas concentrations near the sources.  For example,
a supernova exploding into a dense region of the
interstellar medium~(\cite{Volk}) or into the gas generated by
the strong pre-supernova wind of a massive progenitor star~(\cite{BerezhkoV})
would produce secondary photons
that could show up as point sources.  

Indeed, for many years, observation of $\pi^0$-decay $\gamma$-rays
from the vicinity of shocks around young supernova remnants (SNR)
has been considered a crucial test of the supernova model
of cosmic-ray origin~(\cite{Druryetal}; \cite{Buckley}).  
Note, however, that a sufficiently dense
target is required.  Moreover, it is difficult to distinguish
photons from $\pi^0$ decay from photons originating in 
radiative processes of electrons (~\cite{Petal}).  There are two signatures:
at low energy, observation of a shoulder reflecting the $\pi^0$ peak at
70 MeV would be conclusive.  At higher energy one has to depend
on the hardness and shape of the spectrum for evidence of hadronic 
origin of the photons.  (See \cite{current} for a current review.)

Observation of high-energy neutrinos would be strong evidence for
acceleration of a primary beam of nucleons because such neutrinos
are produced in hadronic interactions.  Expected
fluxes are low~(\cite{GHS}), so large detectors are needed~(\cite{Montaruli,Migneco}).

\subsection{Atmospheric secondaries}
Production of secondary cosmic rays and $\gamma$-rays in the 
interstellar medium generally involves less than one interaction
per primary.  In the language of accelerators, this is the thin-target
regime.  In contrast, the depth of the atmosphere is more than
ten hadronic interaction lengths, so we have a thick target to deal with.
The relevant cascade equation is \vfill\eject
\begin{eqnarray}
{{\rm d}N_i(E,X)\over {\rm d}X}&=&-\left({N_i(E,X)\over\lambda_i(E)}\,
+\,{N_i(E,X)\over d_i(E)}\right)\\ \nonumber
& & +\, \sum_i\int_E^\infty
{N_k(E^\prime,X)\over\lambda_k(E^\prime)}
{F_{k\rightarrow i}(E,E^\prime)\over E}{\rm d}E^\prime,
\label{master}
\end{eqnarray}
where $$F_{k\rightarrow i}\;=\;{1\over\sigma_k}\,E\,
{{\rm d}\sigma_{k\rightarrow i}\over{\rm d}E}.$$
The equation describes the longitudinal 
development of the components of the atmospheric
cascade in terms of slant depth (d$X\,=\,\rho\,$d$\ell$) along 
the direction of the cascade.

The loss terms on the r.h.s. of Eq.~\ref{master} represent interactions and
decay, in analogy to Eq.~\ref{propagation}.  Here 
\begin{equation}
d_i\;=\;\rho\,\gamma\,c\,\tau_i\;=\;\rho\,{E_i\tau_ic\over m_ic^2}
\label{decaylength}
\end{equation}
is the Lorentz dilated decay length of particle $i$ in g/cm$^3$.  The expression
$\lambda_i=d_i$ defines a critical energy below which decay
is more important than re-interaction.  Because the density of
the atmosphere varies with altitude, it is conventional to define
the critical energy at the depth of cascade maximum~({\cite{TKGbook}).
For pions the critical energy in the terrestrial 
atmosphere is $\epsilon_\pi\,=\,115$~GeV,
while $\epsilon_K^\pm=\,850$~GeV.  In astrophysical settings, the
density is usually low enough so that decay always dominates over
hadronic interactions.  An intermediate case of some interest is
production of secondary cosmic rays in the solar chromosphere, where
the scale height is
larger than in the Earth's atmosphere so that decay remains dominant
for another order of magnitude~(\cite{Seckeletal}).

\begin{figure}[htb]
\includegraphics[width=12cm]{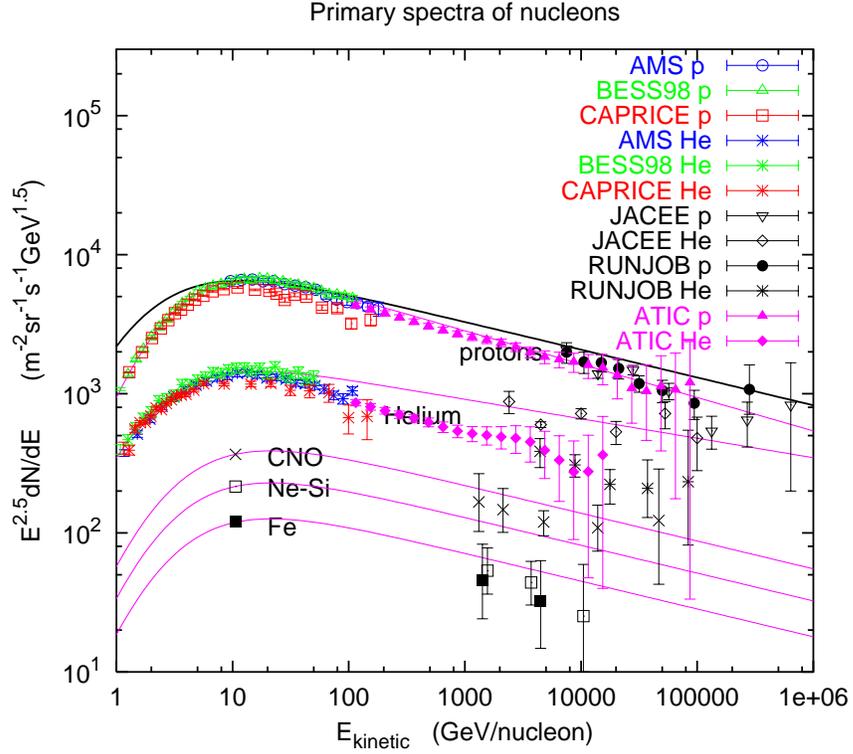}
\caption{The flux of nucleons.  The heavy black line shows the
numerical form of Eq.~\ref{uncorrelated}.  The lighter lines show
extrapolations of fits~(\protect\cite{Hamburg}) to measurements
of protons, helium and three heavier groups below 100 GeV/nucleon.
}
\label{allnucleon_t6}
\end{figure}

The same set of cascade equations (see Eq.~\ref{master}) governs air showers
and uncorrelated fluxes of particles in the atmosphere.  The 
boundary condition for an air shower initiated by a primary of
mass $A$ and total energy $E_0$ is
\begin{equation}
\left.N(X)\right|_{X=0}\,=\,A\,\delta(E\,-\,E_0/A)
\label{EAS}
\end{equation} 
and $N(0)=0$ for all other particles.  This approximation, in
which a nucleus is treated as consisting of independently
interacting nucleons, is
called the superposition approximation.  In practice in
Monte Carlo solutions of the cascade equation it is straightforward
to remove this approximation given a model of nuclear fragmentation, 
(e.g.~\cite{Battistoni}).

The other important boundary condition is that for uncorrelated
fluxes in the atmosphere:
\begin{equation}
\left.N(X)\right|_{X=0}\,=\,\phi_p(E)\;=\;\approx\;1.7\times10^4\,E^{-2.7}\;
({\rm GeV\,cm}^2{\rm s\,sr})^{-1}.
\label{uncorrelated}
\end{equation}
The numerical approximation is for the flux of all nucleons summed
over the five major nuclear groups shown in Fig.~\ref{allnucleon_t6}.
In Eq.~\ref{uncorrelated}, $E$ is total energy per nucleon.  This
numerical approximation is shown as the heavy solid line in 
Fig.~\ref{allnucleon_t6}.  Its curvature at low energy is just
a consequence of plotting the power law in total energy
per nucleon as a function of kinetic
energy per nucleon.  Only a subset of available data is shown
in Fig.~\ref{allnucleon_t6}.  Data from the magnetic spectrometers
BESS98 (\cite{BESS98}) and AMS (\cite{AMSp})
are indistinguishable on the plot for protons, although they differ
somewhat for helium (\cite{AMSHe}).  Data from the
CAPRICE spectrometer (\cite{CAPRICE}) are
15-20\% lower than BESS98 above 10~GeV/nucleon.  The higher energy data
are from balloon-borne emulsion chambers, which are subject to larger
systematic errors because not all energy is sampled in the calorimeter.
Proton and helium data from JACEE (\cite{JACEE}) are shown.
Data of RUNJOB (\cite{RUNJOB1}; \cite{RUNJOB2})
are shown for five groups of nuclei
(protons and helium, CNO, Ne-Si and Fe) above 1000~GeV.
The fits to CNO, Ne-Si and Fe are normalized at 10.6 GeV/nucleon
to data of \cite{HEAO}.

\begin{figure}[htb]
\includegraphics[width=12cm]{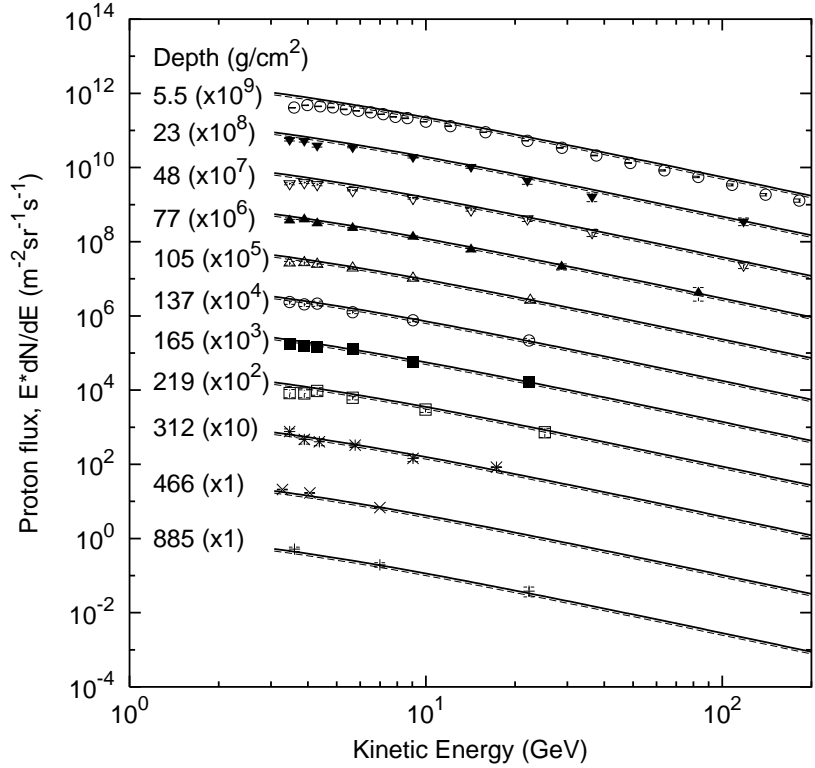}
\caption{The flux of protons.  The data are from the measurements
of~\protect\cite{Mocchiutti} with the CAPRICE detector.
}
\label{patmosphere}
\end{figure}

\subsubsection{Uncorrelated fluxes in the atmosphere}
The simplest physical 
example to illustrate the solution of Eq.~\ref{uncorrelated}
is to calculate the vertical spectrum of nucleons as a function of 
depth in the atmosphere.  Nucleons are stable compared to the
transit time through the atmosphere, so only losses due to 
interactions are important in the cascade equation~\ref{master}.
In the approximation of scaling, the dimensionless distribution
$F_{N\rightarrow N}(E,E^\prime)\rightarrow F(\xi)$ as in Eq.~\ref{scaling},
with $\xi = E/E^\prime$.  Eq.~\ref{master} becomes
\begin{equation}
{{\rm d}N(E,X)\over{\rm d}X}\;=\;-{N(E,X)\over\lambda}
\,+\,{1\over\lambda_N}\,\int_0^1\,N({E\over\xi},X)\,F(\xi){{\rm d}\xi\over\xi^2}.
\label{nucleons}
\end{equation}
The dependence on energy and depth can be factorized, and the solution
of Eq.~\ref{nucleons} is
\begin{equation}
N(E,X)\;=\;K\,e^{-X/\Lambda_N}\times E^{-\alpha},
\label{nucleon2}
\end{equation}
where the boundary condition~\ref{uncorrelated} is satisfied if
$K=1.7 \times 10^4$ and $\alpha = 2.7$.  The attenuation length is
related to the interaction length $\lambda$ by
\begin{equation}
\Lambda_N\;=\;{\lambda_n\over 1\,-\,Z_{NN}},
\label{attenuation}
\end{equation}
where 
$Z_{NN}\,=\,\int_0^1\,\xi^{\alpha -2}\,F_{N\rightarrow N}(\xi)\,{\rm d}\xi\,
\approx\,0.3$
is the spectrum-weighted moment for nucleons, analogous to Eq.~\ref{Zfactor}
for pions.

The  solution outlined above has several obvious approximations such as
neglect of nucleon-anti-nucleon production and neglect of energy-dependence
of the cross sections, but it nevertheless gives a reasonable
representation of measurements of the
spectrum of protons at various atmospheric depths, 
as shown in Fig.~\ref{patmosphere}.  For comparison with measurements
of protons, the solution of Eq.~\ref{nucleon2} for all nucleons
must be modified to remove neutrons, which increase slightly as
a fraction of the total flux of nucleons with increasing depth in the
atmosphere.  The correction, as described in~\cite{TKGbook}, is included
in the calculations shown in Fig.~\ref{patmosphere}.

\begin{figure}[htb]
\includegraphics[width=12cm]{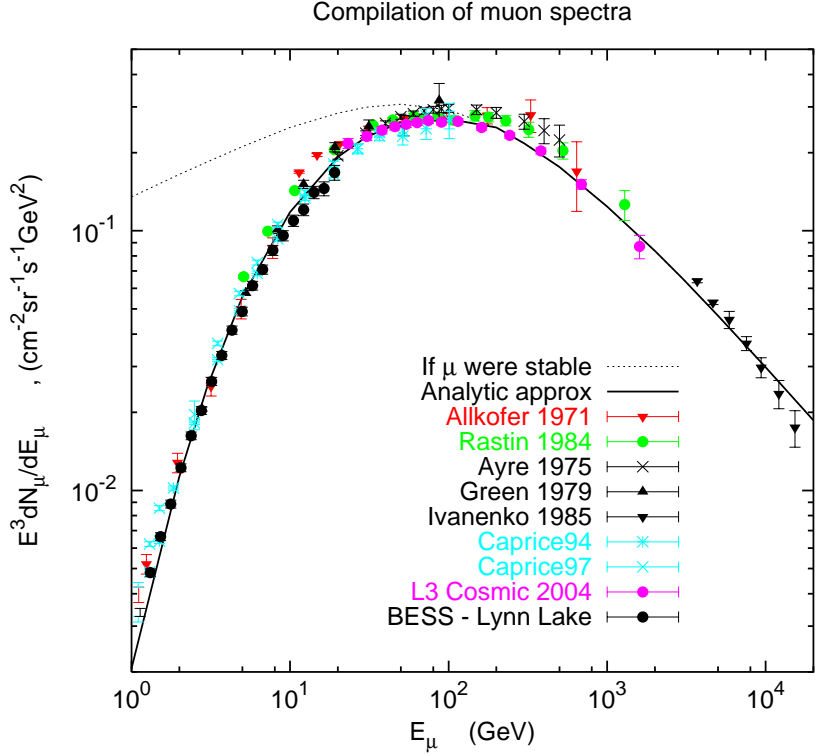}
\caption{Summary of measurements of the vertical muon intensity
at the ground.  The solid line shows an analytic 
calculation~\protect\cite{gaissermu}.
The dotted line shows the spectrum in the absence of decay and
energy loss, or equivalently the muon production spectrum integrated
over the atmosphere.
}
\label{newmuplot}
\end{figure}

Another benchmark measurement of secondary cosmic rays in the
atmosphere is the flux of muons.  The main source of muons is
from decay of charged pions.  There is also a small contribution
from decay of charged kaons, which becomes somewhat more
important at high energy.  At very high energy the muon energy
spectrum  becomes one power steeper than the parent spectrum of nucleons as a
consequence of the extra power of $E_\pi$ in the ratio $\lambda_\pi/d_\pi$,
which represents the decreasing probability of decay relative to
re-interaction for charged pions at high
energy (see Eqs.~\ref{master} and~\ref{decaylength}).  For $E<\epsilon_\pi$
essentially all pions decay, and the muon production spectrum
has the same power behavior as the parent pion and grandparent nucleon
spectrum ($\alpha \approx 2.7$).  At low energy, however,
muon energy-loss and decay become important, and the muon spectrum
at the ground falls increasingly below the production spectrum.
To account for all the complications one generally resorts to 
Monte Carlo calculations.  However, analytic approximations~(\cite{Liparimu}) of these
effects are also possible.  The full line in fig.~\ref{newmuplot} shows
one such calculation~(\cite{gaissermu}), which uses
as input the simple power-law primary spectrum of Eq.~\ref{uncorrelated}.
This simple result compares relatively well to various measurements
in several energy ranges.  The recent data are from CAPRICE (\cite{CAPRICEmu}),
L3-Cosmic (\cite{L3}) and BESS ({\cite{BESSmu}).  One can also see the 
level of relative systematic uncertainties between different measurements.

Although I will not discuss the subject here,
The most important secondary cosmic-ray flux is the atmospheric neutrino
beam because of the discovery of neutrino oscillations
by Super-Kamiokande~(\cite{nudiscovery}).
The experimental situation is reviewed by~\cite{review1} and~\cite{review2}, and
the calculations by~\cite{GH}.

\subsubsection{Air Showers}
Above about 100 TeV cascades generated by individual primary nuclei
have a big enough footprint deep in the atmosphere to trigger an array of widely spaced
detectors on the ground.  The threshold energy may be somewhat lower 
for closely spaced detectors, especially at high altitude.  The
threshold also may be made much higher by separating the detectors
by a large distance.  Examples of the latter are the Akeno Giant Air
Shower Array (AGASA)~(\cite{AGASA}) and the surface
detector of the Auger Project~(\cite{Auger}).  Such ground
arrays work by looking for coincidences in an appropriate time
window, then reconstructing the primary direction and energy from
the timing pattern of the hits and the size of the signals in the 
detectors.  There are large fluctuations from shower to shower
which complicate the interpretation of the data.  The air shower technique
is used at very high energy where the flux is too low to accumulate meaningful
statistics with detectors carried aloft by balloons or spacecraft.
The dividing line at present is approximately $100$~TeV. 

Because of the complicated cascade of interactions that intervenes
between the primary cosmic-ray nucleus incident on the atmosphere
and the sparse data on the ground, Monte Carlo simulations
are used to interpret the data.  The other important reason
for the necessity of a Monte Carlo generation of showers is that
the detectors only sample a tiny fraction of the particles
in the shower.  Simulation of the response of an air shower
detector to showers is therefore crucial.
The standard, fully stochastic, 
four-dimensional air shower generator is CORSIKA (~\cite{Corsika}).
A cascade generator of similar scope and design is AIRES~(\cite{AIRES}).
A fast, one-dimensional cascade generator (\cite{CASC}) that uses libraries of
pre-generated subshowers at intermediate
energies inside cascades is useful for analysis of
ultra-high energy showers measured by fluorescence detectors
for which knowledge of the lateral distribution is less important.
The three-dimensional hybrid generator SENECA~(\cite{SENECA}) uses 
stochastic Monte Carlo methods for the high-energy part of the
shower and at the detector level, but saves time by numerically
integrating the cascade equations for intermediate energies.
The FLUKA program (~\cite{FLUKA1}) is a general code for transport
and interaction of particles through detectors of various types,
including a layered representation of the atmosphere.  The FLUKA interaction
model~(\cite{FLUKA2}) is built into the code and cannot be replaced by a different
event generator.

There is a variety of hadronic event generators on the 
market~(\cite{DPMjet,QGSjet,SIBYLL,VENUS,Nexus}),
which can be called by cascade programs 
like CORSIKA or AIRES to generate showers.
Because the event generators are based on interpolations
between measurements with accelerators at specific points in
phase space and because the energies involved require
extrapolations several orders of magnitude beyond those
accessible with present accelerators, different hadronic 
event generators give different results for observables
in air showers.  We will see examples of this in the discussion
of air showers below.

An air shower detector essentially uses the atmosphere as a calorimeter.
Each shower dissipates a large fraction of its energy as it passes through
the atmosphere, which is sampled in some way by the detector.
It is therefore customary to plot the energy
spectrum in the air shower regime by total energy per particle
rather than by energy per nucleon as at low energy. 
In the lower energy regime the identity
of each primary nucleus can be determined as it passes 
through the detector on a balloon or spacecraft.  
With air showers one has to depend
on Monte Carlo simulations to relate what is measured to the
primary energy and to the mass of the primary particle.  
The resulting energy assignments typically have uncertainties
$\Delta E\,/\,E\,\sim$20-30\%.  Primary mass is often quoted
as an average value for a sample of events in each energy bin,
or at best as a relative fraction of a small number of
groups of elements.

\section{Acceleration}
A detailed review of the theory of particle acceleration by astrophysical
shocks is given in the lectures of~\cite{Ostrowski}.  The main feature necessary
for understanding the implications of air shower data for origin of high-energy
cosmic rays is the concept of maximum energy.  Diffusive, first-order shock
acceleration works by virtue of the fact that particles gain
an amount of energy $\Delta E\,\propto\,E$ at each cycle, where
a cycle consists of a particle passing from the upstream (unshocked)
region to the downstream region and back.  At each cycle, there is
a probability that the particle is lost downstream and does not return
to the shock.  Higher energy particles are those that remain longer
in the vicinity of the shock and have time to achieve high energy.

  After a time $T$ the maximum energy achieved is
\begin{equation}
E_{max}\;\sim\;Ze\beta_s\times B\times T\,V_s,
\label{Emax}
\end{equation}
where $\beta_s = V_s/c$ refers to the velocity of the shock.
This result is an upper limit in that it assumes a minimal diffusion
length equal to the gyroradius of a particle of charge $Ze$
in the magnetic fields behind and ahead of the shock. 
Using numbers typical of Type II supernovae exploding in the average
interstellar medium gives $E_{max}\sim Z\times 100$~TeV (~\cite{Cesarsky}).
More recent
estimates give a maximum energy larger by as much as an order of
magnitude or more for some types of supernovae~(\cite{Berezhko}).

The nuclear charge, $Z$, appears in Eq.~\ref{Emax} because acceleration depends
on the interaction of the particles being accelerated with the moving
magnetic fields.  Particles with the same gyroradius behave in the same way.
Thus the appropriate variable to characterize acceleration is
magnetic rigidity, $R\,=\,pc/Ze\,\approx E_{tot}/Ze$, where $p$ is the total momentum of
the particle.  Diffusive 
propagation also depends on magnetic fields and hence on rigidity.
For both acceleration and propagation, therefore, if there is a
feature characterized by a critical rigidity, $R^*$, then the corresponding
critical energy per particle is $E^*\,=\,Z\times R^*$.  

\begin{figure}[htb]
\includegraphics[width=12cm]{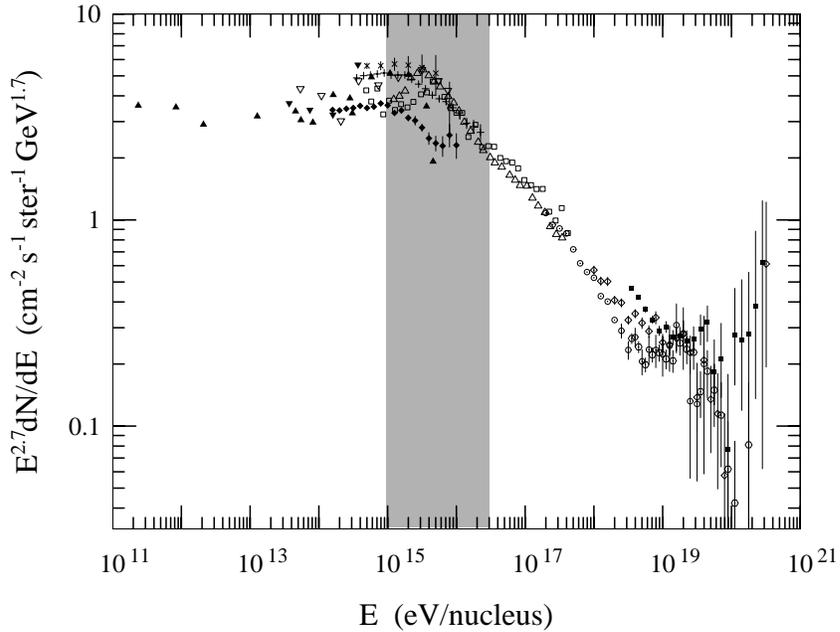}
\caption{High-energy cosmic-ray spectrum.  References to the data
are given in~(\protect\cite{rpp}).  The shaded region indicates a
factor of 30 in total energy (see text).
}
\label{allparticle}
\end{figure}

\section{The anatomy of the cosmic-ray spectrum}

The knee of the spectrum is the steepening that occurs above $10\,^{15}$~eV,
as shown in Fig.~\ref{allparticle}, while the ankle is the 
hardening around $3\times10\,^{18}$~eV.   
One possibility is that the knee
is associated with the upper limit of acceleration by
galactic supernovae, while the ankle is associated
with the onset of an extragalactic population that is less
intense but has a harder spectrum that dominates at sufficiently
high energy.  The general idea that the knee may signal the end
of the population of particles produced in the Galaxy is
an old one that I will trace in the next two subsections.

\subsection{The knee}
If the knee is a consequence of galactic cosmic accelerators reaching
their limiting energy, then there are consequences for energy-dependence
of the composition that can be used to check the idea.  This follows from
the form of Eq.~\ref{Emax}.  Consider first the simplest case in which
all galactic accelerators are identical.  Then $E_{max}\,=\,Z\times R^*$,
where $R^*$ characterizes the maximum rigidity.  When the particles are 
classified by total energy per nucleus, protons will cut off first
at $E_{max}\,=\,e\,R^*$, helium at $E_{max}\,=\,2\,e\,R^*$, etc. 
\cite{Peters} described this cycle of composition change
and pointed out the consequences for composition in a plot like that
reproduced here as Fig.~\ref{Peters}.  Since the abundant elements 
from protons to the iron group cover a factor of 30 in $Z$, the ``Peters cycle''
should occupy a similar range of total energy.  

\begin{figure}[htb]
\includegraphics[width=12cm]{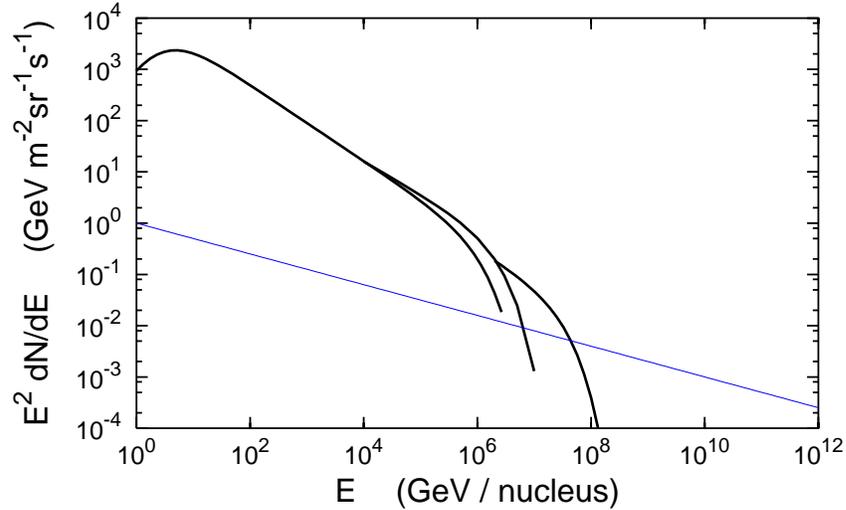}
\caption{Recreation of a figure from~\protect\cite{Peters}
showing a sequence of cutoffs for ions with successively
higher charge when classified by total energy per particle.
The line shows a hypothetical second component with
a harder spectrum than the lower energy component.  (See text
for discussion).
}
\label{Peters}
\end{figure}

Because the observed spectrum does not abruptly stop, Peters hypothesized
a new population of particles coming in with a somewhat harder spectrum,
as indicated by the line in Fig.~\ref{Peters}.  Comparison with the data
in Fig.~\ref{allparticle} shows that reality is more complicated.
The shaded area indicates the factor of 30 for a Peters cycle
assuming $R^*\,=\,10^{15}$~eV.  What is
observed is that the steepened spectrum above
the knee continues on smoothly for more than an order of magnitude in
energy before any sign of a hardening.
Even postulating
a significant contribution from elements heavier than iron 
(up to uranium,~\cite{Jorg})
cannot explain the smooth continuation all the way up to the ankle.

One possibility is that most galactic accelerators cut off
around a rigidity of perhaps $10^{15}$~eV, but a few accelerate
particles to much higher energy and account for the region between
the knee and the ankle~(\cite{AWW}).  This scenario would be 
a generalization of Peters' model.  Its signature
would be a sequence of composition cycles alternating between
light and heavy dominance as the different components from each
source cut off.
 As emphasized by~\cite{Axford}, however,
the problem with this type of model is that
it requires a fine-tuning of the high-energy spectra so that they rise to 
join smoothly at the knee then steepen to fit the data to
$\sim10\,^{18}$~eV.  As a consequence, several models have been proposed 
in which the lower-energy accelerators ($E\,<\,10^{15}$~eV) 
inject seed particles into another process that accelerates 
them to higher energy.  In this way the spectrum above the
knee is naturally continuous with the lower energy region.
One such possibility is acceleration by interaction with
multiple supernova shocks in a cluster of supernovae~(\cite{Axford}).
Another possibility~(\cite{VZ}) is acceleration by a termination shock
in the galactic wind~(\cite{Jokipii}).

The fine tuning problem (i.e. to achieve a smooth spectrum with a sequence of
sources with different maxima) was actually clearly recognized by
\cite{Peters} in his original statement of this idea.  He correctly
pointed out, however, that since the cutoff is a function of rigidity
while the events are classified by a quantity close (but not equal) to total energy,
the underlying discontinuities are smoothed out to some extent.
An interesting question to ask in this context is what power at the source would be 
required to fill in the spectrum from the knee to the ankle.
The answer depends on what is assumed for the spectrum of the sources
and the energy dependence of propagation in this
energy region.  Reasonable assumptions (e.g. $Q(E)\propto E^{-2}$
and $\tau_{esc}\propto E^{-\delta}$ with $\delta\approx 0.3$) lead to an estimate of
$\sim 2\times 10^{39}$~erg/s, less than 10\% of the total power
requirement for all galactic cosmic-rays.  For comparison, the micro-quasar SS433
at 3 kpc distance has a jet power estimated as $10^{39}$~erg/s~(\cite{SS433}).

Another possibility is that the steepening of the spectrum at the knee is
a result of a change in properties of diffusion in the interstellar medium
such that above a certain critical rigidity the characteristic
propagation time $\tau\,_{esc}$ decreases more rapidly with energy.
If the underlying acceleration process were featureless, then
the relative composition as a function of total energy per particle
would change smoothly, with the proton spectrum
steepening first by $0.3$, followed by successively heavier nuclei.
It is interesting that this possibility was also explicitly recognized
by~\cite{Peters}.

A good understanding of the composition would go a long way toward
clarifying what is going on in the knee region and beyond.  
A recent summary of direct measurements of various nuclei shows
no sign of a rigidity-dependent composition change
up to the highest energies 
accessible ($\sim 10\,^{14}$~eV/nucleus) (\cite{Battiston}).
The change associated with the knee is in the air shower regime.
Because of the indirect nature of EAS measurements, however, the composition is
difficult to determine unambiguously.  The composition has to be determined
from measurements of ratios of different components of air showers at
the ground.  For example, a heavy nucleus like iron generates a
shower with a higher ratio
of muons to electrons than a proton shower of the same energy. 
\cite{Swordy} reviewed all available measurements of the
composition at the knee.  A plot of mean log mass ($\langle \ln(A)\rangle$)~(\cite{Kath})
shows no clear pattern when all results are plotted together. 
The best indication at present
comes from the Kascade experiment (~\cite{Kascade}), which shows clear evidence
for a ``Peters cycle'',
the systematic steepening first of hydrogen, 
then of helium, then CNO and finally the 
iron group.  The transition occurs over an energy range 
from approximately $10\,^{15}$~eV
to $3\times10^{16}$~eV, as expected, but the experiment runs out of
statistics by $10^{17}$~eV, so the data do not yet discriminate
among the various possibilities for explaining the spectrum between the knee
and the ankle.

\subsection{The ankle}
Above some sufficiently high energy it seems likely that the cosmic rays
will be of extra-galactic origin.  A proton of energy $10\,^{18}$~eV
has a gyroradius of a kiloparsec in a typical galactic magnetic field,
which is larger than the thickness of the disk of the Galaxy.  
Given constraints from
observed isotropy of particles with $E\sim 10^{19}$~eV, where the
corresponding proton gyroradius is comparable to the full extent
of the Galaxy, the usual assumption is that particles above $10^{19}$
originate outside our galaxy.  There is a suggestion of an anisotropy
just around $10^{18}$~eV from the central regions of
the galaxy (\cite{AGASA18}) that may be due to
neutrons, which survive for a mean pathlength of 
$\sim 10$~kpc at this energy, and could
therefore reach us from the galactic center.
If so, this would suggest that at least some fraction of the cosmic
rays around $10^{18}$~eV are still of galactic origin.  
An interesting discussion of possible implications is given
by~\cite{galcenter}.  In any case,
the questions of whether there is a transition from galactic to extragalactic
cosmic rays and at what energy such a transition occurs are of great
interest.

\begin{figure}[htb]
\includegraphics[width=5.5cm]{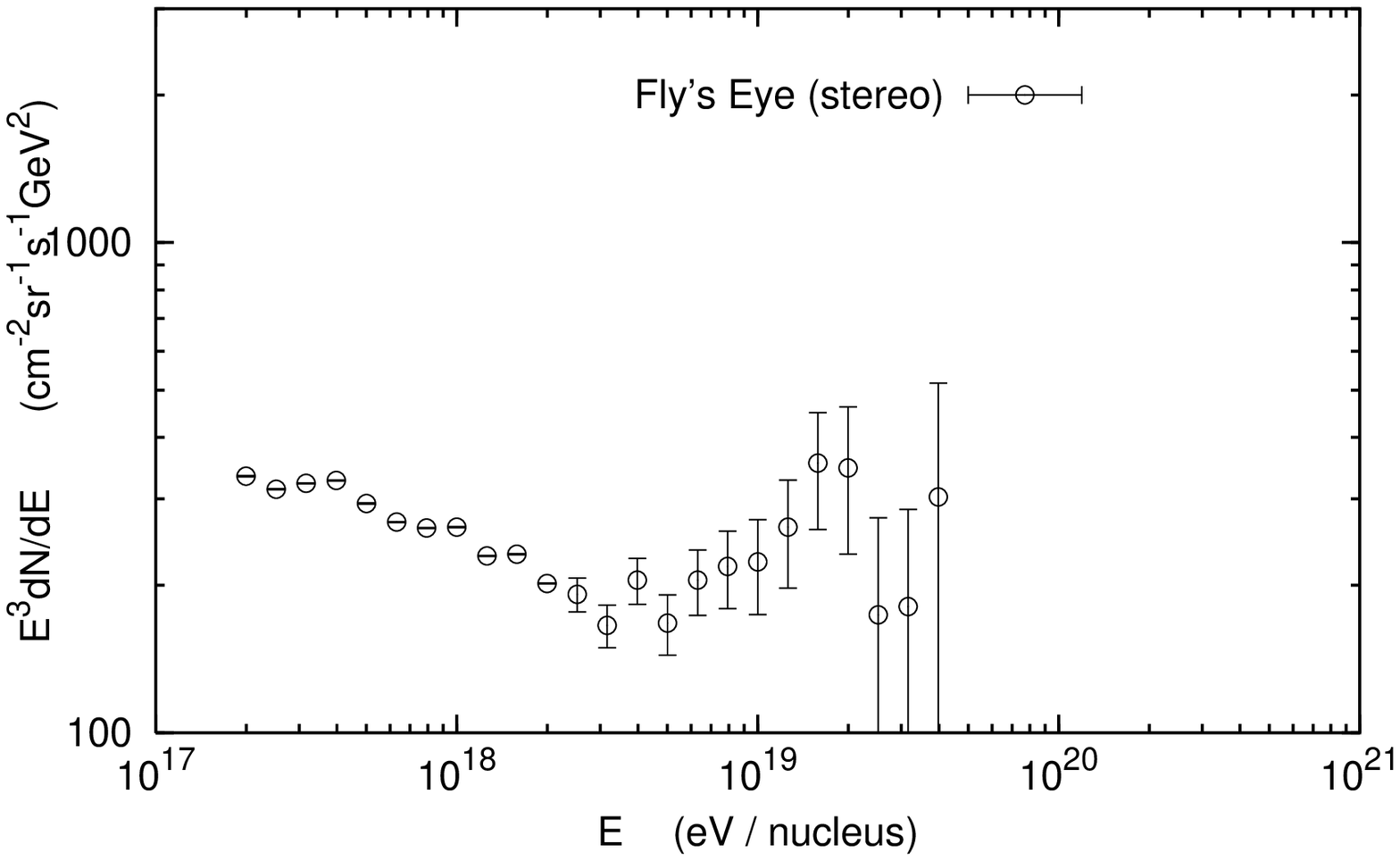}
\includegraphics[width=5.5cm]{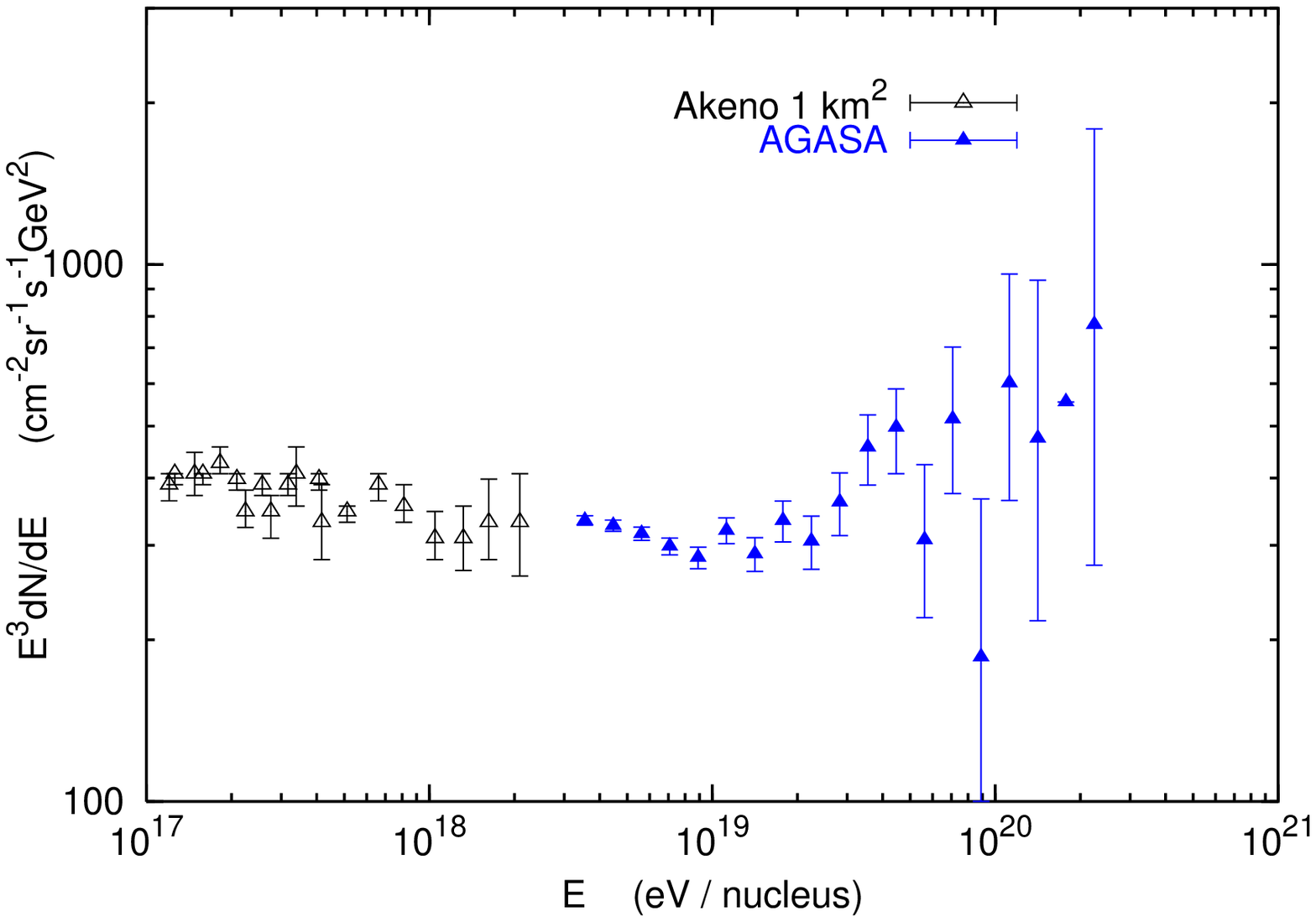}
\includegraphics[width=5.5cm]{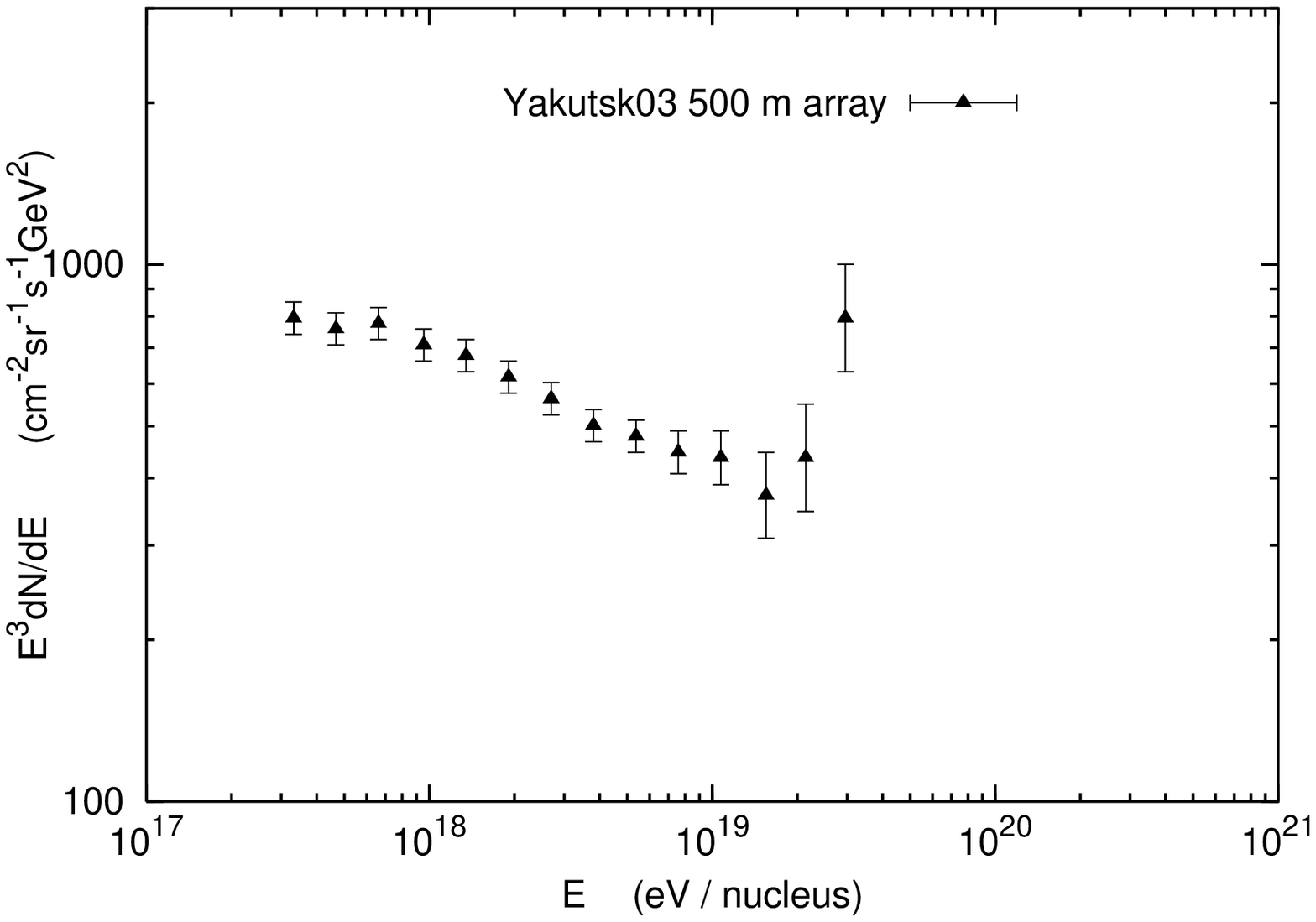}
\includegraphics[width=5.5cm]{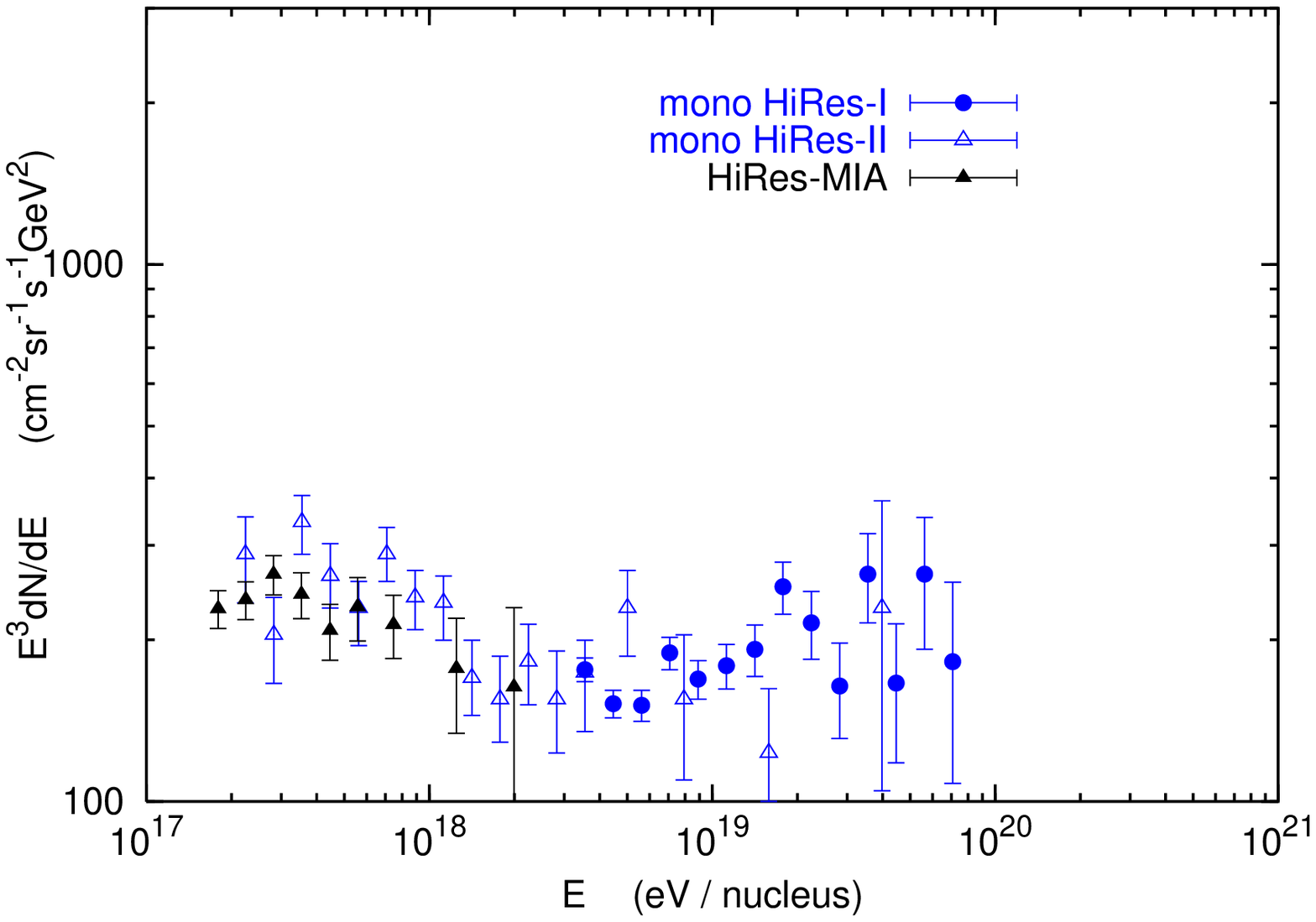}
\caption{Four measurements of the spectrum above $10^{17}~eV$.
}
\label{ankle}
\end{figure}

As background for a discussion of the significance of the ankle and how
it should be interpreted, it is helpful to look at the spectrum 
from individual measurements separately so that systematic differences
between measurements do not obscure any details that may be present. 
Results from four groups are shown in Fig.~\ref{ankle}: 
Fly's Eye stereo (\cite{FlysEyeXmax}); Akeno (\cite{Akeno}) and
AGASA (\cite{AGASA}); Yakutsk (\cite{Yakutsk03}); and the measurements
made with Hi-Res (\cite{HiRes} and \cite{HiResXmax}).  In all these measurements
there is a suggestion of a steepening just below $10^{18}$~eV, which is
sometimes referred to as the ``second knee.''  The ankle appears as
a saddle-like shape with its low point between $3\times10^{18}$ and
$10^{19}$~eV, depending on the experiment.  One could fit the saddle
with a final Peters cycle starting just below $10^{18}$~eV and an
extragalactic component crossing as in
Fig.~\ref{Peters} to contribute to the ankle
in the overall spectrum~(\cite{BahWax}).  Alternatively, it is possible
to make a model in which extragalactic cosmic rays account
for the entire observed flux down to $10\,^{18}$~eV (\cite{BerezinskyGG})
or even lower (~\cite{Bergman}).  The difference lies in the assumptions
made for the spectrum and cosmological evolution of the sources.  I
will return to this issue in the next section.

First it is interesting to ask what data on primary composition may
tell us about the changing populations of particles above $10\,^{17}$~eV.
 In this energy range the composition is measured by the energy
 dependence of the position of shower maximum, $X_{max}$.
An air shower consists of a superposition of electromagnetic cascades
initiated by photons from decay of $\pi\,^0$ particles produced 
by hadronic interactions along the core of the shower as it
passes through the atmosphere.  Most of the energy of the shower
is dissipated by ionization losses of the low-energy electrons and positrons
in these subshowers.  The composite shower reaches a maximum number of
particles (typically $0.7$ particles per GeV of primary energy)
and then decreases as the individual photons fall below the critical
energy for pair production.  Because each nucleus of mass $A$ and total
energy $E_0$ essentially generates $A$ subshowers each of energy $E_0/A$
the depth of maximum depends on $E_0/A$.  Since cascade penetration
increases logarithmically with energy,
\begin{equation}
X_{max}\;=\;\lambda_{ER}\log(E_0/A)\,+\,C,
\label{Xmax}
\end{equation}
where $\lambda_{ER}$ is a parameter (the ``elongation rate'') that depends
on the underlying properties of hadronic interactions in the cascade.

\begin{figure}[thb]
\includegraphics[width=12cm]{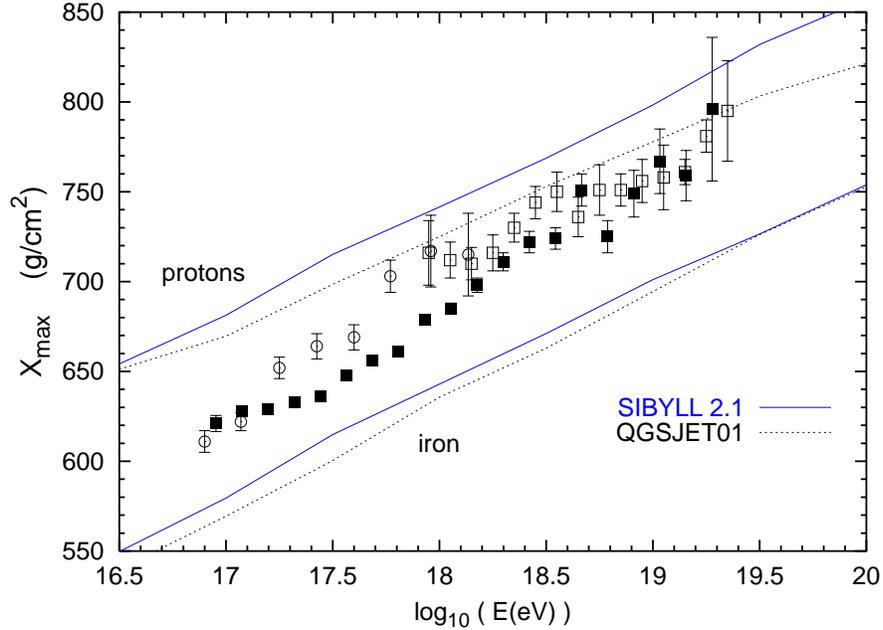}
\caption{Plot of data on mean depth of maximum vs energy.
  Filled squares are data from the stereo Fly's Eye~(\protect\cite{FlysEyeXmax}).
 Open symbols show the data of HiRes (squares)~(\protect\cite{Archbold})
 and HiRes prototype with MIA (circles)~(\protect\cite{HiResXmax}).
This figure also appears in Ref.~(\protect\cite{GaisStan}).
}
\label{xmax}
\end{figure}

Fig.~\ref{xmax} shows results of measurements with the Fly's Eye stereo
detector (\cite{FlysEyeXmax}) compared to measurements of 
HiRes~(\cite{HiResXmax}; \cite{Archbold}).
A weak inference about composition can be made by comparing to the results
of simulations, two of which are shown in the figure.  Both calculations use
CORSIKA (~\cite{Corsika}) with two different interaction models~(\cite{QGSjet,SIBYLL}).
The measurement with the Fly's Eye Stereo detector suggests a transition from a large
fraction of heavies below $10\,^{18}$~eV to a larger fraction of protons by 
$10^{19}$~eV (how much larger depending on which interaction model is chosen).
The coincidence of the change of composition from heavier toward lighter)
 around $\sim 3\times 10^{18}$~eV with the ankle feature in the Fly's Eye
data at the same energy led to the suggestion of a transition from galactic
to extragalactic cosmic rays as a possible explanation~(\cite{FlysEyeXmax}).
This interpretation would favor a model like that of \cite{BahWax}.
The more recent HiRes data set, however, shows the transition from heavier toward
lighter beginning at $10^{17}$~eV and complete by $10^{18}$~eV,
consistent with the models of~\cite{BerezinskyGG} and~\cite{Bergman}.

Because of the uncertainties in the interaction models above accelerator
energies coupled with statistical and systematic limitations of the experiments,
the primary composition as a function of energy above $10^{17}$~eV remains
an open question~(\cite{Watson}).

\section{Highest energy cosmic rays}

Protons lose energy by three loss processes during propagation
through the cosmos.  Red-shift losses (adiabatic losses due to
expansion of the Universe), which apply to all particles, become
important when the distance scales are comparable to the Hubble distance $\approx 4$Gpc.
Protons of sufficiently  high energy also interact with the microwave background
and lose energy to electron-positron pair production 
(for $E\,>\,\sim10^{18}$~eV)
and to photopion production (for $E\,>\,\sim 5\times 10^{19}$~eV).  The
corresponding attenuation lengths (for reducing energy by a factor $1/e$) are
$\lambda_{e^+e^-}\sim 1$~Gpc and $\lambda_{\pi}\sim 15$~Mpc, respectively.
The photo-pion process leads to the expectation of a suppression of the flux
above $5\times10^{19}$~eV unless the sources are within a few tens of Mpc.
The suppression is referred to as the GZK cutoff (or GZK feature) in recognition
of the authors of the two papers,~\cite{Greisen} and~\cite{ZK} 
who first pointed out the effect shortly
after the discovery of the microwave background radiation.

The actual shape of the spectrum at Earth after accounting for these three loss processes
depends on what is assumed 
\begin{itemize}
\item for the spatial distribution of sources, 
\item for the spectrum of accelerated particles at the sources, and 
\item for the possible evolution of activity of the sources on cosmological time scales.  
\end{itemize}
A classic calculation is that of~\cite{Berezinskyetal}, in which the energy-loss
equation is integrated numerically.  This approach neglects effects of fluctuations,
which may be noticeable in certain circumstances.  A recent example of a Monte Carlo
propagation calculation of cosmological propagation is by~\cite{Stanevetal}, which
contains comparisons with other calculations.  Figure~\ref{UHEspectrum}
from~\cite{HiResAGASA} shows an example of a calculated cosmologically evolved
spectrum compared to data of HiRes~(\cite{HiRes}) and AGASA~(\cite{AGASA}).  

\begin{figure}[thb]
\includegraphics[width=12cm]{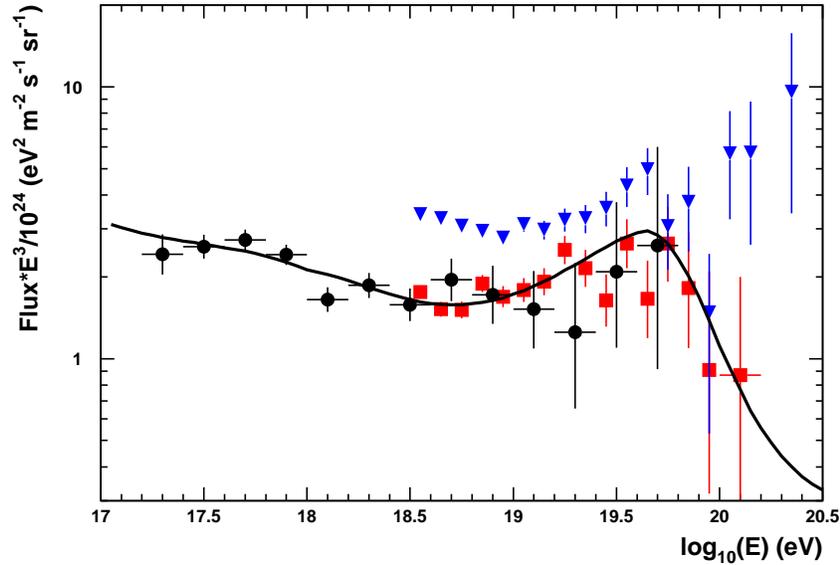}
\caption{The ultra-high-energy cosmic-ray spectrum from a paper
by \protect\cite{HiResAGASA}.  The points with
the fitted curve are from the HiRes fluorescence detectors~(\protect\cite{HiRes}),
while the higher set of points are the AGASA spectrum~(\protect\cite{AGASA}).
See text for a discussion of the curve.
}
\label{UHEspectrum}
\end{figure}

The calculation illustrated in Fig.~\ref{UHEspectrum}
is for a uniform distribution of sources out to large red shifts. 
The ``pile-up'' (which is amplified by plotting $E\,^3\times$~differential flux) is
populated by particles that have fallen below the GZK energy.  The saddle or ankle
feature on the $E\,^3{\rm d}N\,/\,{\rm d}E$ plot
is a consequence of energy losses to pair production.  The data below
$10\,^{18}$~eV have been fit in this example by adding an assumed galactic contribution, which
is not shown separately here (but see~\cite{Bergman} for a discussion).  
As pointed out long ago by~\cite{Hillas1} (see also~\cite{Hillas2}), the degree to
which the extragalactic flux contributes at low energy (for example below $10^{18}$~eV)
can be adjusted by making different assumptions about increased activity 
at large redshift, z $>$ 1.  Such high redshift sources do not contribute at
high energy because of adiabatic losses, but they can contribute at low energy
if they were sufficiently abundant.
On the other hand, it is also possible to construct a model in which the extragalactic
sources only contribute significantly at very high energy (e.g. above 
$10^{19}$~eV), as in~\cite{BahWax}.

The question of the energy above which the extragalactic component accounts for
all of the observed spectrum is important because it is directly related to
the amount of power needed to supply the extragalactic component.  An estimate
of the power needed to supply the extragalactic cosmic rays is obtained by
the replacements $\tau_{esc}\rightarrow\tau_H$ and $\phi(E)\rightarrow\phi_{EG}(E)$
in Eq.~\ref{power}, where $\tau_H\approx 1.4\times 10^{10}$~yrs is the Hubble
time and $\phi_{EG}$ is the extragalactic component of the total cosmic-ray flux.
The result depends on where the EG-component is normalized to the total flux
and on how it is extrapolated to low energy.  A minimal estimate follows from
taking the hardest likely spectral shape (differential index $\alpha = -2$)
and normalizing at $10^{19}$~eV.  This leads to a total power requirement of
$P_{EG}\,=\, 10^{37}$~erg/Mpc$^3$/s.  If the rate of GRBs is 1000 per year,
then they would need to produce $2\times 10^{52}$~erg/burst 
in accelerated cosmic rays to satisfy this
requirement.  The density of AGNs is $\sim 10^{-7}\,/\,{\rm Mpc}^3$~(\cite{Peebles}),
so the corresponding requirement would be $10^{44}$~erg/s per AGN in accelerated
particles.
Normalizing the extragalactic component
at $10^{18}$~eV increases the power requirement by a factor of 10, and assuming
a softer spectral index increases it even more.  These are very rough estimates
just meant to illustrate the relation between the intensity of extragalactic cosmic
rays and the power needed to produce them.  A proper treatment requires an
integration over redshift, accounting for the spatial distribution of sources
and their evolution over cosmic time~(\cite{Stanevbook}).

In addition to being able to satisfy the power requirement, the sources
also  have to be able to accelerate particles to $\sim10\,^{20}$~eV.
The requirements on size and magnetic field in the sources are essentially
given by setting $E_{\rm max}=10^{20}$~eV in Eq.~\ref{Emax} and solving
for the product of $B\times T\,V_s$ (or $B\times R$).  The resulting
constraint on the sources is the famous plot of~\cite{Hillasplot}.
Sites with a sufficiently large $B\times R$ included 
active galactic nuclei (AGN)~(\cite{BerezinskyGG})
and the termination shocks of giant radio galaxies~(\cite{Biermann}).  
Since 1984 two more potential
candidates have emerged, jets of gamma-ray bursts (GRB~\cite{Waxman,Vietri}) 
and magnetars.

Neutrino telescopes may make a decisive contribution to identifying 
ultra-high energy cosmic accelerators by finding point sources of TeV neutrinos
and associating them with known objects, for example with gamma-ray
sources such as AGN or GRB.  Even the detection of a diffuse flux of
high-energy, extraterrestrial neutrinos, or the determination of good
limits can constrain models of cosmic-ray origin under the assumption
that comparable amounts of energy go into cosmic-rays and into neutrinos
produced by the cosmic-rays at the sources (\cite{Anchor}).

As is well-known, the AGASA measurement~(\cite{AGASA}) shows the spectrum
continuing beyond the GZK energy without a suppression, while the HiRes
measurement~(\cite{HiRes}) is consistent with the GZK shape.
The discrepancy between the two experiments is not as great as
it appears in Fig.~\ref{UHEspectrum}.  The difference is amplified quadratically
when the differential spectrum is multiplied by $E^3$ in the plot.
A 20-30\% shift in energy assignment (downward for AGASA or upward for
HiRes) brings the two data sets into agreement
below $5\times 10^{19}$~eV.  A quantitative estimate of
the statistical significance of the difference above $5\times 10^{19}$~eV
is ambiguous because
it depends on what is assumed for the true spectrum.  The difference
is generally considered to be less than three sigma, however~(\cite{Olinto}).
One of the first goals of the Auger project~(\cite{Klages}) is to resolve the question
of whether the spectrum extends significantly beyond the GZK energy.
If not, then higher statistics accumulated by Auger may be used to
constrain models of extra-galactic sources by making precise measurements
both above and below the GZK energy.  If the spectrum does continue
beyond the cutoff then identification of specific, cosmologically nearby sources
from the directions of the ultra-high energy events would be likely.

\section{Conclusion and outlook}
While the questions surrounding the highest energy particles are clearly
of the greatest importance, there are several important open questions
at all energies.  One is the transition, if it exists, from galactic to 
extra-galactic cosmic rays.  Proposed experiments associated with the
telescope array (\cite{TA}) may be optimized to provide good coverage 
down to $10^{17}$~eV (\cite{Thompson}).  

Understanding the knee of the spectrum remains an outstanding problem in
cosmic-ray astrophysics.
Kascade-Grande~(\cite{grande}) 
will cover the energy range from below the knee to $10^{18}$~eV
with a multi-component air shower array at sea level.  The IceCube
detector at the South Pole~(\cite{IceCube}) will have a kilometer square surface 
component, IceTop~(\cite{IceTop}; \cite{IceTop2}),
1.4 km above the top of the cubic kilometer neutrino telescope.
The whole constitutes a novel, large three-dimensional air shower detector
with coverage of the cosmic-ray spectrum from below the knee to $10^{18}$~eV.
In addition to its main mission of neutrino astronomy, IceCube therefore also has
the potential to make important contributions to related cosmic-ray physics.
The high altitude of the surface (9300 m.a.s.l.) allows good primary energy resolution
with the possibility of determining relative importance of the primary
mass groups from the ratio of size at the surface to muon signal in the deep detector.
With a coverage extending to $10^{18}$~eV, these detectors also have the
potential to clarify the location of a transition to cosmic-rays of extragalactic origin.

While the knee of the spectrum remains in the realm of air shower experiments
for the time being, there is much activity aimed at extending direct
measurements of the primary spectrum and composition to reach the knee
from below.  The detectors include
Advanced Thin Ionization 
Calorimeter (ATIC, \cite{ATIC}); Transition Radiation Array for Cosmic Energetic
Radiation (TRACER, \cite{TRACER}); Cosmic Ray Energetics and Mass (CREAM, \cite{CREAM})
and Advanced Cosmic-ray Composition Experiment for Space Station 
(ACCESS, http://www.atic.umd.edu/access.html).
While ACCESS is to be flown in space, ATIC, TRACER and CREAM
all take advantage of NASA's long-duration balloon program which is regularly
achieving flights of two to four weeks in Antarctica.
These experiments also have the opportunity to extend measurements
of the ratio of secondary to primary nuclei to much higher energy
and hence to resolve the related questions about the average source spectral index
and about the isotropy of galactic cosmic rays.


\begin{thebibliography}{99}
\bibitem[{Abassi \textit{ et al.}}, 2002]{HiResAGASA}
{Abassi}, R.U. \textit{ et al.} (2002).
\newblock astro-ph/0208301.
\bibitem[{Abassi \textit{ et al.}}, 2004]{HiRes}
{Abassi}, R.U. \textit{ et al.} (2004).
\newblock \textit{ Phys. Rev. Letters}, {\bf 92}:151101.
\bibitem[{Abu-Zayyad \textit{ et al.}}, 2001]{HiResXmax}
{Abu-Zayyad}, T. \textit{ et al.} (2001).
\newblock \textit{ Ap.J.}, {\bf 557}:686.
\bibitem[{Achard \textit{ et al.}}, 2004]{L3}
{Achard}, P. \textit{ et al.} (2004).
\newblock \textit{ Phys. Lett. B}, {\bf 598}:15.
\bibitem[{Ahrens \textit{ et al.}}, 2004a]{IceCube}
{Ahrens}, J. \textit{ et al.} (2004a).
\newblock \textit{ Astropart. Phys.}, {\bf 20}:507.
\bibitem[{Ahrens \textit{ et al.}}, 2004b]{Kath}
{Ahrens}, J. \textit{ et al.} (2004b).
\newblock \textit{ Astropart. Phys.}, {\bf 21}:565.
\bibitem [{Alcarez \textit{ et al.}}, 2000a]{AMSp}
{Alcarez}, J. \textit{ et al.} (2000a)
\newblock \textit{ Phys. Lett. B}, {\bf 490}:27.
\bibitem [{Alcarez \textit{ et al.}}, 2000b]{AMSHe}
{Alcarez}, J. \textit{ et al.} (2000b)
\newblock \textit{ Phys. Lett. B}, {\bf 494}:193.
\bibitem[{Alvarez-Mu\~{n}iz \textit{ et al.}}, 2002]{CASC}
{Alvarez-Mu\~{n}iz}, J. \textit{ et al.} (2002).
\newblock \textit{ Phys. Rev. D}, {\bf 66}:033011.
\bibitem[{Allkofer {et al.}}, 1971]{Allkofer}
{Allkofer}, O.C., {Carstensen}, K. \& {Dau}, W.D. (1971).
\newblock \textit{ Phys. Lett. B}, {\bf 36}:425.
\bibitem[{Anchordoqui \textit{ et al.}}, 2004a]{galcenter}
{Anchordoqui}, L.A., {Goldberg}, H., {Halzen}, F. \& {Weiler}, T.J. (2004).
\newblock \textit{ Phys. Lett. B}, {\bf 593}:42.
\bibitem[{Anchordoqui \textit{ et al.}}, 2004b]{Anchor}
{Anchordoqui}, L.A., {Goldberg}, H., {Halzen}, F. \& {Weiler}, T.J. (2004).
\newblock hep-ph/0410003.
\bibitem[{Apanasenko \textit{ et al.}}, 2001]{RUNJOB1}
{Apanasenko}, A.V. \textit{ et al.} (2001).
\newblock \textit{ Astropart. Phys.}, {\bf 16}:13.
\bibitem[{Arai \textit{ et al.}}, 2003]{TA}
{Arai}, Y. \textit{ et al.} (2003).
\newblock in \textit{ Proc. 28th Int. Cosmic Ray Conf.} (Tsukuba, ed.
T. Kajita \textit{ et al.}, Universal Academy Press), {\bf 2}:1025. 
\bibitem[{Archbold \& Sokolsky}, 2003]{Archbold}
{Archbold}, G. \& {Sokolsky}, P. (2003).
\newblock in \textit{ Proc. 28th Int. Cosmic Ray Conf.} (ed.
T. Kajita \textit{ et al.}, Universal Academy Press) {\bf 1}:405.
\bibitem[{Asakimori \textit{ et al.}}, 1998]{JACEE}
{Asakimori}, K. \textit{ et al.} (1998).
\newblock \textit{ Ap.J.}, {\bf 502}:278.
\bibitem[{Axford}, 1994]{Axford}
{Axford}, W.I. (1994).
\newblock \textit{ Ap.J. Suppl.}, {\bf 90}:937.
\bibitem[{Ayre \textit{ et al.}}, 1975]{Ayre}
{Ayre}, C.A. \textit{ et al.} (1975).
\newblock \textit{ J. Phys. G}, {\bf 1}:584.
\bibitem[{Bahcall \& Waxman}, 2003]{BahWax}
{Bahcall}, J.N. \& {Waxman}, E. (2003).
\textit{ Phys. Lett. B}, {\bf 556}:1.
\bibitem[{Battiston}, 2004]{Battiston}
{Battiston}, R. (2004).
\newblock in \textit{ Frontiers of Cosmic Ray Science} (ed.
T. Kajita \textit{ et al.}, Universal Academy Press):229.
\bibitem[{Battistoni, \textit{ et al.}}, 1997]{Battistoni}
{Battistoni}, G., {Forti}, C., {Ranft}, J. \& {Roesler}, S. (1997).
\newblock \textit{ Astropart. Phys.}, {\bf 7}:49.
\bibitem[{Battistoni \textit{ et al.}}, 2004]{FLUKA2}
{Battistoni}, G. \textit{ et al.} (2004).
\newblock astro-ph/0412178
\bibitem[{Berezhko}, 1996]{Berezhko}
{Berezhko}, E.G. (1996).
\newblock \textit{ Astropart. Phys.}, {\bf 5}:367.
\bibitem[{Berezhko \& Ellison}, 1999]{nonlinear} 
{Berezhko}, E.G. \& {Ellison}, D.C. (1999).
\newblock \textit{ Ap.J.}, {\bf 526}:385.
\bibitem[{Berezhko \& V\"{o}lk}, 2000]{Volk}
{Berezhko}, E.G. \& {V\"{o}lk}, H.J. (2000).
\newblock \textit{ Astropart. Phys.}, {\bf 14}:201.
\bibitem[{Berezhko, P\"{u}hlhofer \& V\"{o}lk}, 2003]{BerezhkoV}
{Berezhko}, E.G., {P\"{u}hlhofer}, G. \& {V\"{o}lk}, H.J. (2003).
\newblock \textit{ Astron. Astrophys.}, {\bf 400}:971.
\bibitem[{Berezinsky \& Grigorieva}, 1988]{Berezinskyetal}
{Berezinsky}, V.S. \& {Grigorieva}, S.I. (1988).
\newblock \textit{ Astron. Astrophys.}, {\bf 199}:1.
\bibitem[{Berezinsky \textit{ et al.}}, 2004]{BerezinskyGG}
 {Berezinsky}, V., {Gazizov}, A. \& {Grigorieva}, S.
\newblock astro-ph/0410650.
\bibitem[{Bergman}, 2004]{Bergman}
{Bergman}, D. (2004).
\newblock astro-ph/0407244.
\bibitem[{Biermann \& Strittmatter}, 1987]{Biermann}
{Biermann}, P. \& {Strittmatter}, P.A. (1987).
\newblock \textit{ Ap.J.}, {\bf 322}:643.
\bibitem[{Bird, \textit{ et al.}}, 1993]{FlysEyeXmax}
{Bird}, D.J. \textit{ et al.} (1993).
\newblock \textit{ Phys. Rev. Letters}, {\bf 71}:3401.
\bibitem[{Boezio \textit{ et al.}}, 1999]{CAPRICE}
{Boezio}, M \textit{ et al.} (1999).
\newblock \textit{ Ap.J.}, {\bf 518}:457.
\bibitem[{Bopp \textit{ et al.}}, 2004]{DPMjet}
{Bopp}, F.W., {Ranft}, J., {Engel},, R. \& {Roesler}, S. (2004).
\newblock astro-ph/0410027 and references therein.
\bibitem[{Bossard \textit{ et al.}}, 2001]{Nexus}
{Bossard}, G. \textit{ et al.} (2001).
\newblock \textit{ Phys. Rev. D}, {\bf 63}:054030.
\bibitem[{Bottino \textit{ et al.}}, 1998]{exotica}
{Bottino}, A. \textit{ et al.} (1998).
\newblock \textit{ Phys. Rev. D}, {\bf 58}:123503.
\bibitem[{Buckley \textit{ et al.}}, 1998]{Buckley}
{Buckley}, J.H. \textit{ et al.} (1998).
\newblock \textit{ Astron. Astrophys.}, {\bf 329}:639.
\bibitem[{Distefano \textit{ et al.}}, 2002]{SS433}
{Distefano}, C., {Guetta}, D., {Waxman}, E. \& {Levinson}, A.
\newblock \textit{ Ap.J.}, {\bf 575}:378.
\bibitem[{Drescher \& Farrar}, 2003]{SENECA}
{Drescher}, H.-J. \& {Farrar}, G. (2003).
\newblock \textit{ Phys. Rev. D}, {\bf 67}:116001.
\bibitem[{Drury \textit{ et al.}}, 1994]{Druryetal}
{Drury}, L.O'C., {Aharonian}, F.A. \& {V\"{o}lk}, H.J. (1994).
\newblock \textit{ Astron. Astrophys.}, {\bf 287}:959.
\bibitem[{Drury \textit{ et al.}}, 2001]{current}
{Drury}, L.O'C. \textit{ et al.} (2001).
\newblock \textit{ Ap. Sci. Rev.}, {\bf 99}:329.
\bibitem[{DuVernois \textit{ et al.}}, 2001]{HEAT}
{DuVernois}, M.A. \textit{ et al.} (2001).
\newblock \textit{ Ap.J.}, {\bf 559}:296.
\bibitem[{Engel \textit{ et al.}}, 2001]{SIBYLL} 
{Engel}, R., {Gaisser}, T.K. \& {Stanev}, Todor (2001).
\newblock \textit{ Proc. 27th
Int. Cosmic Ray Conf.} (ed. K.-H. Kampert, G. Heinzelmann \& C. Spiering,
Copernicus Gesellschaft, Hamburg) {\bf 2}:431.
\bibitem[{Engelmann \textit{ et al.}}, 1990]{HEAO}
{Engelmann}, J.J. \textit{ et al.} (1990).
\newblock \textit{ Astron. Astrophys.}, {\bf 233}:96.
\bibitem[{Erlykin \& Wolfendale}, 2001]{AWW}
{Erlykin}, A.D. \& {Wolfendale}, A.W. (2001).
\newblock \textit{ J. Phys. G}, {\bf 27}:1005.
\bibitem[{Fass\`{o} \textit{ et al.}}, 2001]{FLUKA1}
{Fass\`{o}}, A., {Ferrari}, A., {Ranft}, J. \& {Sala}, P.R. (2001).
\newblock in \textit{ Proc. MonteCarlo 2000, Lisbon} (ed. A. Kling \textit{ et al.},
Springer-Verlag, Berlin):955.
\bibitem[{Fukuda \textit{ et al.}}, 1998]{nudiscovery}
{Fukuda}, Y. \textit{ et al.} (1998).
\newblock \textit{ Phys. Rev. Letters}, {\bf 81}:1562.
\bibitem[{Furukawa \textit{ et al.}}, 2003]{RUNJOB2}
{Furukawa}, M. \textit{ et al.} (2003).
\newblock in \textit{ Proc. 28th Int. Cosmic Ray Conf.} (Tsukuba, ed.
T. Kajita \textit{ et al.}, Universal Academy Press), {\bf 4}:1837.
\bibitem[{Gaisser}, 1990]{TKGbook}
{Gaisser}, T.K. (1990).
\newblock \textit{ Cosmic Rays and Particle Physics}
(Cambridge University Press).
\bibitem[{Gaisser, Halzen \& Stanev}, 1995]{GHS}
{Gaisser}, T.K., {Halzen}, F. \& {Stanev}, Todor (1995).
\newblock \textit{ Phys. Reports}, {\bf 258}:173.
\bibitem[{Gaisser, Protheroe \& Stanev}, 1998]{Petal}
{Gaisser}, T.K. {Protheroe}, R.J. \& {Stanev}, Todor (1998).
\newblock \textit{ Ap.J.}, {\bf 492}:219.
\bibitem[{Gaisser, Honda, Lipari \& Stanev}, 2001]{Hamburg}
{Gaisser}, T.K., {Honda}, M., {Lipari}, P. \& {Stanev}, Todor (2001).
\newblock \textit{ Proc. 27th Int. Cosmic Ray Conf.} (Hamburg) {\bf 5}:1643.
\bibitem[{Gaisser}, 2002]{gaissermu}
{Gaisser}, T.K. (2002).
\newblock \textit{ Astropart. Phys.}, {\bf 16}:285.
\bibitem[{Gaisser \& Honda}, 2002]{GH}
{Gaisser}, T.K. \& {Honda}, M. (2002).
\textit{ Ann. Revs. Nucl. Part. Sci.}, {\bf 52}:153.
\bibitem[{Gaisser \textit{ et al.}}, 2003]{IceTop}
{Gaisser}, T.K. \textit{ et al.} (2003).
\newblock in \textit{ Proc. 28th Int. Cosmic Ray Conf.} (Tsukuba,
ed. T. Kajita \textit{ et al.}, Universal Academy Press), {\bf 2}:1117.
\bibitem[{Gaisser \& Stanev}, 2004]{rpp}
{Gaisser}, T.K. \& {Stanev}, Todor (2004).
\newblock in S. Eidelman \textit{ et al., Reviews of Particle
Properties, Physics Letters B}, {\bf 592}:228.
\bibitem[{Gaisser \& Stanev}, 2005]{GaisStan}
{Gaisser}, T.K. \& {Stanev}, Todor (2005).
\newblock \textit{ Nucl. Phys. A}:(to be published).
\bibitem[{Glushkov \textit{ et al.}}, 2003]{Yakutsk03}
{Glushkov}, A.V. \textit{ et al.} (2003).
\newblock in \textit{ Proc. 28th Int. Cosmic Ray Conf.} (Tsukuba, ed.
T. Kajita \textit{ et al.}, Universal Academy Press), {\bf 1}:389.
\bibitem[{Goodman}, 2005]{Goodman}
{Goodman}, Jordan (2005).
\newblock This volume.
\bibitem[{Green}, 1979]{Green}
{Green}, P.J. \textit{ et al.} (1979).
\newblock \textit{ Phys. Rev. D}, {\bf 20}:1598.
\bibitem[{Greisen}, 1966]{Greisen}
{Greisen}, K. (1966).
\textit{ Phys. Rev. Letters}, {\bf 16}:748.
\bibitem[{Hayashida \textit{ et al.}}, 1999]{AGASA18}
{Hayashida}, N. \textit{ et al.} (1999).
\newblock \textit{ Astropart. Phys.}, {\bf 10}:303.
\bibitem[{Heck \& Knapp}, 2003]{Corsika}
{Heck}, D. \& {Knapp}, J. (2003).
\newblock \textit{ Extensive Air Shower simulation with CORSIKA: A user's Guide},
(V 6.020, FZK Report, February 18, 2003).\\
\newblock http://www-ik.fzk.de/corsika/
\bibitem[{Heinbach \& Simon}, 1995]{reacceleration2}
{Heinbach}, U. \& {Simon}, M. (1995).
\newblock \textit{ Ap.J.}, {\bf 441}:209.
\bibitem[{Hillas}, 1968]{Hillas1}
{Hillas}, A.M. (1968).
\newblock \textit{ Canadian J. Phys.}, {\bf 46}:S623.
\bibitem[{Hillas}, 1974]{Hillas2}
{Hillas}, A.M. (1974).
\newblock \textit{ Phil. Trans. R. Soc. Lond. A}, {\bf 277}:413. 
\bibitem[{Hillas}, 1984]{Hillasplot}
{Hillas}, A.M. (1984).
\textit{ Ann. Revs. Astron. Astrophys.}, {\bf 22}:425.
\bibitem[{H\"{o}randel}, 2004]{Jorg}
{H\"{o}randel}, J.R. (2004).
\newblock \textit{ Astropart. Phys.}, {\bf 21}:241.
\bibitem[{Huangs \textit{ et al.}}, 2003]{grande}
{Huangs}, A. \textit{ et al.} (2003).
\newblock in \textit{ Proc. 28th Int. Cosmic Ray Conf.} (Tsukuba, ed.
T. Kajita \textit{ et al.}, Universal Academy Press), {\bf 2}:985.
\bibitem[{Hunter \textit{ et al.}}, 1997]{Egret}
{Hunter}, S.D. \textit{ et al.} (1997).
\newblock \textit{ Ap.J.}, {\bf 481}:205.
\bibitem[{Ivanenko \textit{ et al.}}, 1985]{Ivanenko}
{Ivanenko}, I.P. \textit{ et al.} (1985).
\newblock in \textit{ Proc. 19th Int. Cosmic Ray Conf., La Jolla} 
(NASA Conf. Publ. No 2376) {\bf 8}:21
\bibitem[{Jokipii \& Morfill}, 1987]{Jokipii}
{Jokipii}, J.R. \& {Morfill}, G.E. (1987).
\newblock \textit{ Ap.J.}, {\bf 312}:170.
\bibitem[{Jones \textit{ et al.}}, 2001]{Jonesetal}
{Jones}, F.C., {Lukasiak}, A., {Ptuskin}, V. \& {Webber}, W. (2001).
\newblock \textit{ Ap.J.}, {\bf 547}:246.
\bibitem[{Jung \textit{ et al.}}, 2001]{review2}
{Jung}, C.K., {Kajita}, T., {Mann}, T. \& {McGrew}, C. (2001).
\newblock \textit{ Ann. Rev. Nucl. Part. Sci.}, {\bf 51}:451.
\bibitem[{Kajita \& Totsuka}, 2001]{review1}
{Kajita}, T. \& {Totsuka}, Y. (2001).
\newblock \textit{ Revs. Mod. Phys.}, {\bf 73}:85.
\bibitem[{Kalmykov \textit{ et al.}}, 1997]{QGSjet}
{Kalmykov}, N.N., {Ostapchenko}, S.S. \& {Pavlov}, A.I. (1997).
\newblock \textit{ Nucl. Phys. B (Proc. Suppl.)}, {\bf 52}:17.
\bibitem[{Klages}, 2005]{Klages}
{Klages}, H. (2005).
\newblock This volume.
\bibitem[{Kremer \textit{ et al.}}, 1999]{CAPRICEmu}
{Kremer}, J. \textit{ et al.} (1999).
\newblock \textit{ Phys. Rev. Letters}, {\bf 83}:4241.
\bibitem[{Lagage \& Cesarsky}, 1983]{Cesarsky}
{Lagage}, P.O. \& {Cesarsky}, C.J. (1983).
\newblock \textit{ Astron. Astrophys.}, {\bf 118}:223 and {\bf 125}:249.
\bibitem[{Lipari}, 1993]{Liparimu}
{Lipari}, P. (1993).
\newblock \textit{ Astropart. Phys.}, {\bf 1}:195. 
\bibitem[{Migneco}, 2005]{Migneco}
{Migneco}, E. (2005).
\newblock This volume.
\bibitem[{Mocchiutti}, 2003]{Mocchiutti}
{Mocchiutti}, E. (2003).
\newblock \textit{ Atmospheric and Interstellar Cosmic Rays
Measured with the CAPRICE98 Experiment} (Thesis, KTH, Stockholm)
\bibitem[{Montaruli}, 2003]{Montaruli}
{Montaruli}, T. (2003).
\newblock astro-ph/0312558 (\textit{ Nucl. Phys. B (Suppl.)}, to be published).
\bibitem[{Moskalenko \& Strong}, 1998]{MoskStronge}
{Moskalenko}, I.V. \& {Strong}, A.W. (1998).
\newblock \textit{ Ap.J.}, {\bf 493}:694.
\bibitem[{Moskalenko \textit{ et al.}}, 1998]{MoskStrongpbar}
{Moskalenko}, I.V., {Strong}, A.W., {Ormes}, J.F. \& 
{Potgieter}, M.S. (2002).
\newblock \textit{ Ap.J.}, {\bf 565}:280.
\bibitem[{Motoki \textit{ et al.}}, 2003]{BESSmu}
{Motoki}, M. \textit{ et al.} (2003).
\newblock \textit{ Astropart. Phys.}, {\bf 19}:113.
\bibitem[{M\"{u}ller}, 2005]{TRACER}
{M\"{u}ller}, D. (2005).
\newblock This volume and http://tracer.uchicago.edu
\bibitem[{Nagano \textit{ et al.}}, 1992]{Akeno}
{Nagano}, M. \textit{ et al.} (1992).
\newblock \textit{ J. Phys. G}, {\bf 18}:423.
\bibitem[{Olinto}, 2004]{Olinto}
{Olinto}, A. (2004).
\newblock in \textit{ Frontiers of Cosmic Ray Science} (ed.
T. Kajita \textit{ et al.}, Universal Academy Press):299.
\bibitem[{Orito \textit{ et al.}}, 2000]{BESSpbar}
{Orito}, S. \textit{ et al.} (2000).
\newblock \textit{ Phys. Rev. Letters}, {\bf 84}:1078.
\bibitem[{Ostrowski}, 2005]{Ostrowski}
{Ostrowski}, M. (2005).
\newblock This volume.
\bibitem[{Peebles}, 1993]{Peebles}
{Peebles}, P.J.E. (1993).
\newblock \textit{ Principles of Physical Cosmology} (Princeton University Press).
\bibitem[{Peters}, 1961]{Peters}
{Peters}, B. (1961).
\newblock \textit{ Nuovo Cimento}, {\bf XXII}:800.
\bibitem[{Rastin}, 1984]{Rastin}
{Rastin}, R.C. (1984).
\newblock \textit{ J. Phys. G}, {\bf 10}:1609.
\bibitem[{Roth}, 2003]{Kascade}
{Roth}, M. \textit{ et al.} (2003).
\newblock in \textit{ Proc. 28th Int. Cosmic Ray Conf.} (Tsukuba, ed.
T. Kajita \textit{ et al.}, Universal Academy Press), {\bf 1}:139.
\bibitem[{Sanuki, \textit{ et al.}}, 2000]{BESS98}
{Sanuki}, T. \textit{ et al.} (2000).
\newblock \textit{ Ap.J.}, {\bf 545}:1135.
\bibitem[{Sciutto}, 2001]{AIRES}
{Sciutto}, S.J. (2001).
\newblock \textit{ Proc. 27th
Int. Cosmic Ray Conf.} (ed. K.-H. Kampert, G. Heinzelmann \& C. Spiering,
Copernicus Gesellschaft, Hamburg) {\bf 1}:237.\\
\newblock http://www.fisica.unlp.edu.ar/auger/aires/
\bibitem[{Seckel, Stanev \& Gaisser}, 1991]{Seckeletal}
{Seckel}, D. {Stanev} Todor \& {Gaisser}, T.K. (1991).
\newblock \textit{ Ap.J.}, {\bf 382}:651.
\bibitem[{Seo \& Ptuskin}, 1994]{reacceleration1}
{Seo}, E.S. \& {Ptuskin}, V.S. (1994).
\newblock \textit{ Ap.J.}, {\bf 431}:705.
\bibitem[{Seo \textit{ et al.}}, 2003]{CREAM}
{Seo}, E.-S. \textit{ et al.} (2003).
\newblock in \textit{ Proc. 28th Int. Cosmic Ray Conf.}
(Tsukuba, ed. T. Kajita \textit{ et al.}, Universal Academy Press), {\bf 4}:2101.
\newblock http://cosmicray.umd.edu/cream/cream.html
\bibitem[{Shapiro \& Silberberg}, 1970]{ShapSilb} 
{Shapiro}, Maurice M. \& {Silberberg}, Rein (1970).
\newblock \textit{ Ann. Revs. Nucl. Sci.}, {\bf 20}:323.
\bibitem[{Stanev \textit{ et al.}}, 2000]{Stanevetal}
{Stanev}, Todor \textit{ et al.} (2000).
\textit{ Phys. Rev. D}, {\bf 62}:093005.
\bibitem[{Stanev}, 2003]{Stanevbook}
{Stanev}, Todor (2003).
\newblock \textit{ High Energy Cosmic Rays} (Springer Verlag, Berlin). 
\bibitem[{Stanev, \textit{ et al.}}, 2005]{IceTop2}
{Stanev}, T. for the IceCube Collaboration (2005).
\newblock astro-ph/0501046.
\bibitem[{Stecker}, 1971]{Stecker}
{Stecker}, F.W. (1971).
\newblock \textit{ Cosmic Gamma Rays} (NASA Scientific and Technical Information
Office, NASA SP-249).
\bibitem[{Swordy \textit{ et al.}}, 2002]{Swordy}
{Swordy}, S. \textit{ et al.} (2002).
\newblock \textit{ Astropart. Phys.}, {\bf 18}:129.
\bibitem[{Takeda \textit{ et al.}}, 2003]{AGASA} 
{Takeda}, M. \textit{ et al.} (2003).
\newblock \textit{ Astropart. Phys.}, {\bf 19}:447.
\bibitem[{Thompson}, 2004]{Thompson}
{Thompson}, G. (2004).
\newblock Talk given at Leeds Workshop on Ultra-High Energy Cosmic Rays,
22 July.
\bibitem[{Vietri}, 1995]{Vietri}
{Vietri}, M. (1995).
\newblock \textit{ Ap.J.}, {\bf 453}:883.
\bibitem[{V\"{o}lk \& Zirakashvili}, 2003]{VZ}
{V\"{o}lk}, H.J. \& {Zirakashvili}, V.N. (2003).
\newblock in \textit{ Proc. 28th Int. Cosmic Ray Conf.}
(Tsukuba, ed. T. Kajita \textit{ et al.}, Universal Academy Press), {\bf 4}:2031.
\bibitem[{Watson \textit{ et al.}}, 2004]{Auger}
{Watson}, A.A. \textit{ et al.} (2004).
\newblock {Nucl. Inst. Methods A}, {\bf 523}:50.
\bibitem[{Watson}, 2004]{Watson}
{Watson}, A.A. (2004).
\newblock astro-ph/0410514.
\bibitem[{Waxman}, 1995]{Waxman}
{Waxman}, E. (1995).
\newblock \textit{ Phys. Rev. Letters}, {\bf 75}:386.
\bibitem[{Wefel \textit{ et al.}}, 2003]{ATIC}
{Wefel}, John \textit{ et al.} (2003).
\newblock in \textit{ Proc. 28th Int. Cosmic Ray Conf.}
(Tsukuba, ed. T. Kajita \textit{ et al.}, Universal Academy Press), {\bf 4}:1849.
\bibitem[{Wefel}, 2005]{Wefel} 
{Wefel}, John (2005).
\newblock This volume.
\bibitem[{Werner}, 1993]{VENUS}
{Werner}, K. (1993).
\newblock \textit{ Phys. Rep.}, {\bf 232}:87.
\bibitem[{Zatsepin \& Kuz'min}, 1966]{ZK}
{Zatsepin}, G.T. \& {Kuz'min}, V.A. (1966).
\textit{JETP Letters}, {\bf 4}:78.

\end{thebibliography}
\end{document}